\documentclass[journal]{IEEEtran}
\RequirePackage[T1]{fontenc}

\usepackage{standalone}
\usepackage{cite}
\usepackage{amsmath,amsfonts,amsthm}
\usepackage{dsfont}
\usepackage{siunitx}
\usepackage{graphicx}
\usepackage[caption=false,font=footnotesize,labelfont=sf,textfont=sf]{subfig}
\usepackage{tikz}
\usepackage{circuitikz}
\usetikzlibrary{positioning, arrows.meta, calc, fit}
\usepackage{pgfplots}
\pgfplotsset{compat=newest}
\usepackage{array}
\usepackage{tabularx}
\usepackage{booktabs}
\usepackage{multirow}
\usepackage[normal, flushleft]{threeparttable}
\usepackage{algpseudocode}
\usepackage{algorithm}
\usepackage{xcolor}
\usepackage{hyperref}
\hypersetup{
    colorlinks,
    linkcolor={red!30!black},
    citecolor={blue!30!black},
    urlcolor={blue!80!black}
}
\usepackage{style/macros}
\usepackage{style/catppuccinpalette}

\usepackage[table, dvipsnames]{xcolor}

\newtheorem{remark}{Remark}

\title{Learning Optimal Power Flow with \\ Pointwise Constraints}

\author{
Damian Owerko$^\dagger$, 
Anna Scaglione$^\ddagger$, 
and Alejandro Ribeiro$^\dagger$
\thanks{\noindent $^\dagger$Department of Electrical \& Systems Engineering, University of Pennsylvania. $^\ddagger$Department of Electrical \& Computer Engineering, Cornell University.}}

\date{June 30, 2025}

\begin{document}
\bstctlcite{IEEEexample:BSTcontrol}

\maketitle

\begin{abstract}

Training learning parameterizations to solve optimal power flow (OPF) with pointwise constraints is proposed. In this novel training approach, a learning parameterization is substituted directly into an OPF problem with constraints required to hold over \emph{all} problem instances. This is different from existing supervised learning methods in which constraints are required to hold across the \emph{average} of problem instances. Training with pointwise constraints is undertaken in the dual domain with the use of augmented Lagrangian and dual gradient ascent algorithm. Numerical experiments demonstrate that training with pointwise constraints produces solutions with smaller constraint violations. Experiments further demonstrated that pointwise constraints are most effective at reducing constraint violations in corner cases -- defined as those realizations in which constraints are most difficult to satisfy. Gains are most pronounced in power systems with large numbers of buses. 

\end{abstract}

\section{Introduction}
Optimal power flow (OPF) is a critical optimization problem for the energy industry used to allocate electricity generation throughout the day, establish day-ahead market prices and plan for grid infrastructure \cite{Cain12-HistoryOptimalPower}. The objective of OPF is to minimize the cost of generating electrical power, subject to constraints imposed by the grid infrastructure, physical laws and demand patterns (Section \ref{ch4:sec:opf}). Due to the sinusoidal nature of alternating current and the constraints imposed by electrical interconnections, OPF is a non-convex optimization problem \cite{Lesieutre05-ConvexitySet}, which is known to be NP-hard~\cite{Bienstock19-StrongNPhardness}. The difficulty of solving OPF is exacerbated by the increased penetration of renewable energy, battery storage and electric vehicles. These technologies increase peak loads and load variability~\cite{Rizvi18-ElectricVehiclesTheir}, both of which make OPF more difficult to solve~\cite{Rossi24-OptimalPowerFlow}. 

This paper is a contribution to the use of machine learning to accelerate computation of OPF solutions (see Section \ref{sec_related_work} for context). In particular, our goal is to learn parameterizations that estimate solutions of OPF with small constraint violations. This goal is well motivated because existing approaches tend to learn solutions with large constraint violations. This is a significant drawback because constraints do not only represent user requirements but also codify the laws of physics. Solutions that violate constraints by significant margins are \emph{not} true representations of realizable system states.

To learn parameterizations that estimate solutions of OPF with better feasibility profiles we adopt a supervised learning approach in which the learning parameterization is substituted directly into the OPF problem of interest (Section \ref{ch4:sec:primal-dual}). We observe that while most of the supervised learning literature formulates problems with \emph{expected} constraints [cf. \eqref{eqn_opf_learning_average}] this is not the right formulation to guarantee constraint satisfaction across all realizations. Rather, we must use learning with \emph{pointwise} constraints whereby we require that constraints be satisfied for each individual realization [cf. \eqref{eqn_opf_learning_pointwise_not_augmented}].

To solve pointwise constrained learning problems we operate in the dual domain. We do so with the use of an augmented Lagrangian (Section \ref{sec_duality}) and a block coordinate stochastic dual gradient ascent algorithm (Section \ref{sec_dual_training}). The unique aspect of this algorithm relative to existing supervised learning approaches is that we use separate Lagrange multipliers for each separate realization of power demand. To make this clear we describe (standard) unsupervised learning approaches with average constraints and show that they can be interpreted as learning with pointwise constraints if all realizations share the same Lagrange multipliers (Section \ref{sec_shared_multipliers}). 

We undertake numerical experiments to evaluate the merits of learning with pointwise constraints (Section \ref{ch4:sec:experiments}). These experiments are done on the IEEE 30, IEEE 57, IEEE 118, GOC 179, and IEEE 300 power systems \cite[ver. 21.07]{Babaeinejadsarookolaee21-Power} and use graph attention networks (\ref{ch4:sec:gat}) for their learning parameterizations. We compare pointwise constraints to the use of shared multipliers (average constraints) and supervised learning (\ref{ch4:sec:supervised}) approaches to OPF. These experiments corroborate the following hypothesis (Section \ref{sec_feasibility_experiments}):

\begin{list}{}{
\setlength{\labelwidth}{18pt}%
\setlength{\labelsep}{2pt}%
\setlength{\leftmargin}{20pt}%
\setlength{\topsep}{5pt}
\setlength{\itemsep}{5pt}}

\item[\textbf{(H1)}] Because training with pointwise constraints enforces constraints on individual realizations it results in learned parameterizations with smaller constraint violations. Sometimes, but not always, improved feasibility comes at the cost of a reduction in optimality (Figures \ref{ch4:tab:model_summary}, \ref{ch4:tab:case_summary} and \ref{ch4:tab:constraint_breakdown}). 

\item[\textbf{(H2)}] Because larger power systems with more buses have more constraints, reductions in constraint violations are most significant in larger networks. Indeed, we see little difference in IEEE 30 and IEEE 57 but reductions of up to 4 times in relative constraint violations in IEEE 118, GOC 179, and IEEE 300 (Figures \ref{ch4:tab:case_summary} and \ref{ch4:tab:constraint_breakdown}).

\item[\textbf{(H3)}] Because pointwise constraints accentuate infrequent but large deviations, constraint violations are most reduced in corner cases -- i.e., realizations in which constraints are most difficult to satisfy. Indeed, constraint violations in the scatter plots in Figures \ref{ch4:fig:tradeoff_mean} and \ref{ch4:fig:tradeoff_max} are not only shifted towards smaller violations but are also more concentrated. 

\end{list}

\noindent We conclude from (H1)-(H3) that standard supervised and unsupervised learning methods for OPF problems are flawed. Learning with pointwise constraints mitigates these flaws.


\subsection {Related Work} \label{sec_related_work}

The importance and difficulty of solving OPF has motivated the development of heuristics and relaxations. These include linear~\cite{Sun06-DCOptimalPower, Chatzivasileiadis18-Lecture}, quadratic~\cite{Hijazi17-ConvexQuadratic, Sundar19-OptimizationBased}, conic~\cite{Jabr06-RadialDistributionLoad} and semi-definite programming relaxations~\cite{Bai08-SemidefiniteProgramming, Jabr12-ExploitingSparsitySDP}. These relaxations have varying degrees of accuracy in different regimes but it has been generally demonstrated that they produce near-optimal solutions \cite{Molzahn16-ConvexRelaxations, Low14-ConvexRelaxationOptimal, Kocuk16-StrongSOCPRelaxations}. Coupled with the development of computational frameworks for OPF~\cite{Coffrin18-PowerModelsjl}, it is fair to state that we can solve OPF problems at reasonable computational cost. In, e.g., the Grid Optimization Competition~\cite{Petra21-SolvingRealistic, Aravena23-RecentDevelopments, Curtis21-Decomposition, Holzer24-GOCompetition}, teams have solved realistic grid scale security constrained OPF problems in less than 45-minutes~\cite{Petra21-SolvingRealistic}. 

Such remarkable progress notwithstanding, the computational costs of the interior point methods \cite{Castillo13-Computational} or the iterative linear programming algorithms \cite{Castillo16-SuccessiveLinear} used to solve OPF relaxations are still substantial. This motivates a long running effort to leverage machine learning to further reduce the cost of solving OPF~\cite{AlRashidi09-Applications}. Approaches to learn solutions of OPF can be supervised (\ref{ch4:sec:supervised}) or unsupervised (Section \ref{ch4:sec:primal-dual}). 

In supervised methods a learning parameterization is trained to imitate the output of a traditional solver \cite{Guha09-MachineLearningAC,Owerko20-OptimalPowerFlow}. Supervised approaches often produce solutions with large constraint violations. A technique proposed to mitigate this problem is the addition of constraint violation penalties to the imitation cost \cite{Huang22-DeepOPFVSolvingACOPF, Pan21-DeepOPFDeepNeural, Pan23-DeepOPF, Zhou23-DeepOPFFTOneDeep}. However, penalties have limited effect on constraint violations \cite{Yoo21-DynamicPenaltyFunction, Wang21-Understanding, Huang24-UnsupervisedLearning} as they tend to yield problems that are sensitive to hyperparameter choice \cite{Yoo21-DynamicPenaltyFunction} and often exhibit gradient pathologies that prevent training~\cite{Wang21-Understanding, Huang24-UnsupervisedLearning}. 

Unsupervised learning approaches substitute a learning parameterization directly into OPF formulations~\cite{Wang23-FastOptimalPower, Owerko24-UnsupervisedOptimal}. To solve the resulting optimization problem the use of fixed~\cite{Donti20-DC3LearningMethod} and adaptive~\cite{Huang21-DeepOPFNGTFast, Huang24-UnsupervisedLearning, Wang21-Understanding} penalties has been proposed. More principled works propose the use of Lagrangian relaxations of OPF, which can be solved using dual gradient ascent~\cite{Yan20-RealtimeOptimalPower, Fioretto20-PredictingAC, Fioretto21-LagrangianDuality, Chen22-UnsupervisedDeep, Calvo-Fullana23-State, Kim25-UnsupervisedDeep, Giraud24-ConstraintdrivenDeep, Yuan24-UnsupervisedLearning, Yan22-HybridDatadrivenMethod}. Unsupervised learning in the Lagrangian dual domain accounts for constraints explicitly -- if in expectation -- and for that reason it is not surprising that they are best at yielding solutions with small feasibility gap~\cite{Yan20-RealtimeOptimalPower, Fioretto20-PredictingAC, Fioretto21-LagrangianDuality, Chen22-UnsupervisedDeep, Calvo-Fullana23-State, Kim25-UnsupervisedDeep, Giraud24-ConstraintdrivenDeep, Yuan24-UnsupervisedLearning, Yan22-HybridDatadrivenMethod}. 

This paper is unique in the OPF literature for its use of pointwise constraints. 
We point out that although this may seem like a minor modification it is in actuality a very different approach to learning with constraints. It is akin to the difference between expectations of random variables -- which is how costs are defined in regular constrained learning -- and their distributions -- which are shaped by the imposition of pointwise constraints. Training with pointwise constraints also makes a substantial difference in the reduction of constraint violations. This is most marked in corner cases (Figures \ref{ch4:fig:tradeoff_mean} and \ref{ch4:fig:tradeoff_max}) and is in line with observations in the foundational papers of constrained learning theory \cite{pacc, feasible_learning}. At the time of writing, such pointwise formulations have only been studied in the context of reinforcement learning for dynamical optimal power flow \cite{Wu24-ConstrainedReinforcement, Su24-ReviewSafeReinforcement}. In these works, the goal is to find a control policy that maintains feasibility as the load is gradually perturbed. This is distinct from the problem considered here.

\nocite{Baker22-EmulatingACOPF, Wang21-FeasibilityACDC}



\section{Optimal Power Flow}\label{ch4:sec:opf}

%


\begin{figure}
    
\centering
    
{

\small

\begin{tikzpicture}[european]
    
    \draw
    (0,0) node[below]{$v_i$} to[oosourcetrans, *-, a={$t_{ij}:1$}, i>={$f_{ij}$}]
    ++(3,0) coordinate (A) to[generic, l={$y_{ij}$}]
    ++(3,0) coordinate (B) to[short, -*, i<={$f_{ji}$}]
    ++(1.0,0) node[below]{$v_j$};
    
    \draw (A) to[generic, *-, l={$y^c_{ij}$}] ++(0,-1.5) node[tground]{};
    
    \draw (B) to[generic, *-, l={$y^c_{ji}$}] ++(0,-1.5) node[tground]{};

\end{tikzpicture}

}
    
\caption{Circuit Diagram of a $\pi$-Section Branch Model with Transformer. A branch $(i,j)$ from bus $i$ to bus $j$ is shown. Variables $v_i \in \complex$ and $v_j \in \complex$ denote voltages at buses $i$ and $j$, whereas variables $f_{ij}\in \complex$ and $f_{ji}\in \complex$ represent power flowing from bus $i$ to bus $j$ and to bus $i$ from bus $j$, respectively. The branch is characterized by a transformer ratio $t_{nm} \in \complex$, line admittance $y_{ij} \in\complex$, and shunt admittances $y^{\text{C}}_{ij} \in \complex$ and $y^{\text{C}}_{ji} \in \complex$.}
    
\label{ch4:fig:branch}

\end{figure}
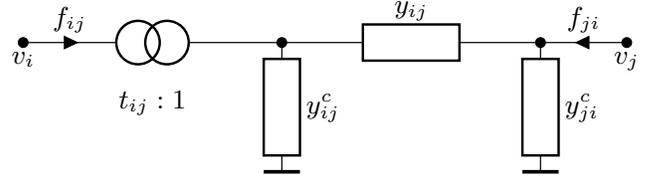


We model an electrical grid as a network of $N$ \emph{buses} connected by a set $E$ containing $M$ directed \emph{branches}. Power is produced or consumed at buses and flows between them via branches \cite{Chatzivasileiadis18-Lecture}. Associated with each bus $i$, we have complex voltage variables $v_i \in \bbC$ and associated with each branch $(i,j)$ we have variables $f_{ij}\in \bbC$ representing power flows from node $i$ to node $j$ and variables $f_{ji}\in \bbC$ representing power flows from node $j$ to node $i$; see Figure \ref{ch4:fig:branch}. We adopt a generalized \(\pi\)-section branch model~\cite{Zimmerman11-MATPOWER, Coffrin18-PowerModelsjl} in which a branch $(i,j)$ is characterized by a transformer ratio $t_{ij} \in \bbC$, shunt admittances $y^{\text{C}}_{ij}\in \bbC$ and $y^{\text{C}}_{ji}\in \bbC$ on each side of the branch and a line admittance $y_{ij}\in\bbC$. With this characterization of a branch, the power flows $f_{ij}$ and $f_{ji}$ satisfy,
\begin{alignat}{4}
    f_{ij} 
        & ~=~ \Big(\, y_{ij} + y^{\text{C}}_{ij} \,\Big)^* \, 
                  \left| \frac{v_{i}}{t_{ij}}  \right|^2
                      & ~-~ y_{ij}^* \,\frac{\,v_i \, v_j^*\,}{t_{ij}} ~,
                                \label{eqn_branch_from_elementwise} \\
    f_{ji} 
        & ~=~ \Big(\, y_{ij} + y^{\text{C}}_{ji} \, \Big)^* \, \,
                  |\hspace{2.4pt} v_{j} \hspace{2.4pt}|^2  
                      & ~-~ y_{ij}^* \, \frac{\,v_i^* \, v_j\,}{t_{ij}^*} ~,
                                \label{eqn_branch_to_elementwise}
\end{alignat}
where $x^*$ represents the complex conjugate of variable $x\in\bbC$. Observe that buses $(i,j)$ are directed and this directionality is important because \eqref{eqn_branch_from_elementwise} and \eqref{eqn_branch_to_elementwise} are not symmetric expressions. Further note that $f_{ij}$ and $f_{ji}$ are both representing power flowing into the branch. These flows are \emph{not} opposites of each other in general. 

We further associate shunt admittances $y^{\text{S}}_{i}\in \bbC$ with each bus along with a variable $s_i \in \bbC$ to denote power generated at bus $i$ and a variable $r_i \in \bbC$ to represent power demand at bus $i$. Power flows, power generation, power demand, and voltages at any bus $i$ are related by Kirchoff's current law. If we let $n(i)$ denote the set of buses $j$ that share a branch $(i,j)$ or $(j,i)$ with bus $i$, we write Kirchoff's current law as,
\begin{equation}\label{eqn_power_balance}
    s_i - r_i 
        ~=~ \sum_{j\in n(i)} f_{ij}
                ~-~ y_{i}^{\text{S}*} \, |v_i|^2  .
\end{equation}
We read \eqref{eqn_power_balance} as stating that the power $\sum_{j\in n(i)} f_{ij}$ flowing out of bus $i$ is the difference between the powers $s_i$ and $r_i$ generated and consumed (demanded) at the bus. The term $y_{i}^{\text{S}*} |v_i|^2$ represents power lost at bus $i$.


\begin{figure}

\centering

\addtolength{\jot}{0.5em}

\hrule

{

\small

\begin{alignat}{3}
    &\textbf{Input} \quad 
        && \bfr
               \nonumber
                   \\
    &\underset{\bfs, \bfv}{\textbf{Minimize}} \qquad 
        &&  \sum_{i=1}^{N} 
               c_{0i} + c_{1i} \Re(s_i) + c_{2i} \Re^2(s_i) 
                   \tag{\ref{eqn_cost}} 
                       \\ 
    &\textbf{Subject to} \qquad
        && s_i - r_i 
              ~=~ \sum_{j\in n(i)} f_{ij}
                      ~-~ y_{i}^{\text{S}*} \, |v_i|^2 
                          \tag{\ref{eqn_power_balance}} 
                              \\
   &   && s^{\text{G}}_{i, \min} \leq s_i \leq s^{\text{G}}_{i,\max} 
                          \tag{\ref{eqn_generator_limit}} 
                              \\
   &    && v_{i,\min} \leq |v_i| \leq v_{i, \max} ,
                          \tag{\ref{eqn_voltage_limit}} 
                              \\
   &    && |f_{ij}| \leq f_{ij, \max},  \quad |f_{ji}| \leq f_{ji, \max} 
                          \tag{\ref{eqn_power_limit}} 
                              \\
   &    && \theta_{ij,\min} \leq \angle (v_i v_j^*)  \leq \theta_{ij, \max} 
                          \tag{\ref{eqn_voltage_angle_difference}} 
                              \\
    &\textbf{With} \quad
        && f_{ij} = \Big(\, y_{ij} + y^{\text{C}}_{ij} \,\Big)^* \, 
                         \left| \frac{v_{i}}{t_{ij}}  \right|^2
                             - y_{ij}^* \,\frac{\,v_i \, v_j^*\,}{t_{ij}} 
                                 \tag{\ref{eqn_branch_from_elementwise}} 
                                     \\
    &   && f_{ji} = \Big(\, y_{ij} + y^{\text{C}}_{ji} \, \Big)^* \, \,
                         |\hspace{2.4pt} v_{j} \hspace{2.4pt}|^2  
                             - y_{ij}^* \, \frac{\,v_i^* \, v_j\,}{t_{ij}^*}  \qquad
                                 \tag{\ref{eqn_branch_to_elementwise}} 
                                     \\
    &\textbf{Output} \quad
        && \bfs^{\dagger} (\bfr), \bfv^\dagger (\bfr)
              \nonumber 
\end{alignat}    

}

\hrule
    
\caption{Optimal Power Flow. We seek optimal power generation $\bfs^{\dagger}(\bfr)$ and voltage $\bfv^\dagger(\bfr)$ values that minimize the cost $C(\bfs)$ [cf. \eqref{eqn_cost}] to meet demand $\bfr$ while satisfying electrical circuit equations [cf. \eqref{eqn_branch_from_elementwise}, \eqref{eqn_branch_to_elementwise} and \eqref{eqn_power_balance}] as well as physical limits on voltages and power flows [cf. \eqref{eqn_generator_limit}, \eqref{eqn_voltage_limit}, \eqref{eqn_power_limit} and \eqref{eqn_voltage_angle_difference}].}

\label{fig_OPF}
        
\end{figure}


Given an electrical grid -- as characterized by buses $i$ with shunt admittance $y_{i}^{\text{S}}$ and branches $(i,j)$ with transformer ratios $t_{ij}$, shunt admittances $y^{\text{C}}_{ij}$ and $y^{\text{C}}_{ji}$ and line admittances $y_{ij}$ -- voltage, power and flow variables that satisfy \eqref{eqn_branch_from_elementwise}, \eqref{eqn_branch_to_elementwise} and \eqref{eqn_power_balance} are physically realizable in theory. In practice, we have several other physical limits. The first of these are lower and upper bounds on the power generated at each bus,
\begin{equation}\label{eqn_generator_limit}
    s^{\text{G}}_{i, \min} \leq s_i \leq s^{\text{G}}_{i,\max}  .
\end{equation}
The inequalities here are between complex numbers and are meant to hold separately in the real and imaginary parts. They represent capacity limits on power plants. A second restriction pertains to minimum and maximum allowable voltages at each bus, which is expressed on the voltage's absolute value,
\begin{equation}\label{eqn_voltage_limit}
    v_{i,\min} \leq |v_i| \leq v_{i, \max} .
\end{equation}
The goal of this constraint is to maintain the voltage at each bus within a margin of a reference operational value. The third constraint is on the power flowing into each branch. This constraint is imposed for all branches $(i, j)$ and is on the absolute value of the flows coming from each bus, 
\begin{align}\label{eqn_power_limit}
    |f_{ij}| \leq f_{ij, \max},  \quad
    |f_{ji}| \leq f_{ji, \max} . 
\end{align}
These constraints model the physical limit of power that can flow in a transmission line. The fourth requirement is that the phases of the voltages at the ends of each branch must be similar. We impose this requirement by constraining the phase $\angle (v_i v_j^*)$ of the voltage product $v_i v_j^*$ to a target range
\begin{align}\label{eqn_voltage_angle_difference}
    \theta_{ij,\min} \leq \angle (v_i v_j^*)  \leq \theta_{ij, \max} .
\end{align}
The purpose of this constraint is to improve the stability of the transmission system \cite{Cain12-HistoryOptimalPower}.

Voltage, power and flow variables that satisfy the electrical power flow relationships in  \eqref{eqn_branch_from_elementwise}, \eqref{eqn_branch_to_elementwise} and \eqref{eqn_power_balance}, as well as the operating range constraints in \eqref{eqn_generator_limit}, \eqref{eqn_voltage_limit},
\eqref{eqn_power_limit} and \eqref{eqn_voltage_angle_difference} are feasible in an electrical grid specified by the admittances and transformer gains that appear in \eqref{eqn_branch_from_elementwise}, \eqref{eqn_branch_to_elementwise} and \eqref{eqn_power_balance} and the operating range bounds that appear in \eqref{eqn_generator_limit}, \eqref{eqn_voltage_limit},
\eqref{eqn_power_limit} and \eqref{eqn_voltage_angle_difference}. In optimal power flow (OPF) we are given demand requirements $\bfr=[s_1^{\text{D}}; \ldots; s_N^{\text{D}}]$ and are tasked with finding generation powers $\bfs=[s_1^{\text{G}}; \ldots; s_N^{\text{G}}]$ that optimize some metric of generation cost. In this paper we assume that generation cost is measured by a quadratic function of the form
\begin{equation}\label{eqn_cost}
    C(\bfs) 
        = \sum_{i=1}^{N} 
              c_{0i} + c_{1i} \Re(s_i) + c_{2i} \Re^2(s_i).
\end{equation}
Thus, OPF is defined here as the optimization program in which we want to find variables $\bfs$ that minimize the cost $C(\bfs)$ in \eqref{eqn_cost} subject to satisfying the constraints in \eqref{eqn_branch_from_elementwise}, \eqref{eqn_branch_to_elementwise}, \eqref{eqn_power_balance}, \eqref{eqn_generator_limit}, \eqref{eqn_voltage_limit},
\eqref{eqn_power_limit} and \eqref{eqn_voltage_angle_difference}. To reduce complexity of the optimization program, flow variables $f_{ij}$ and $f_{ji}$ are substituted for their values in \eqref{eqn_branch_from_elementwise} and \eqref{eqn_branch_to_elementwise}. Thus, the problem we want to solve in this paper is the minimization of the cost $C(\bfs)$ in \eqref{eqn_cost} subject to satisfying the constraints in \eqref{eqn_power_balance}, \eqref{eqn_generator_limit}, \eqref{eqn_voltage_limit},
\eqref{eqn_power_limit} and \eqref{eqn_voltage_angle_difference} with $f_{ij}$ and $f_{ji}$ substituted by their values in \eqref{eqn_branch_from_elementwise}, \eqref{eqn_branch_to_elementwise}. We use $\bfs^{\dagger}(\bfr)$ and $\bfv^\dagger(\bfr)$ to denote (optimal) power and voltage variables that solve OPF. This OPF problem formulation is summarized in Figure \ref{fig_OPF} for future reference.

We also summarize in Figure \ref{fig_variables} the set of variables that appear in Figure \ref{fig_OPF}. Power demand vectors $\bfr = [s^{\text{D}}_1;\ldots;s^{\text{D}}_N]$ are inputs to the OPF problem in Figure \ref{fig_OPF} whereas power generations $\bfs =  [s^{\text{G}}_1;\ldots;s^{\text{G}}_N]$ and voltages $\bfv=[v_1;\ldots;v_N]$ are variables. These three types of variables are associated with individual buses. There are $N$ of each type -- we set $r_i = 0$ or $s_i = 0$ to represent buses without demand or generation capacity. Flow variables $f_{ij}$ and $f_{ji}$ are associated with each branch. There are $M$ of them and they are implicit in Figure \ref{fig_OPF} since we choose to substitute them for their explicit expressions in \eqref{eqn_branch_from_elementwise} and \eqref{eqn_branch_to_elementwise}.

Figure \ref{fig_parameters} summarizes all the parameters that appear in Figure \ref{fig_OPF}. Some of these parameters are associated with buses and some others with branches. As we will reference them to build graph attention networks in \ref{ch4:sec:gat} we define here bus and branch feature vectors. The bus feature vector $\bfw_i$ collects all of the parameters associated with bus $i$,
\begin{equation}\label{eqn_bus_features}
    \bfw_i :=
    [
    y_{i}^{\text{S}};
    s^{\text{G}}_{i, \min}; s^{\text{G}}_{i,\max};
    v_{i,\min}, v_{i, \max};
    c_{0i}; c_{1i}, c_{2i}
    ]
\end{equation}
This feature includes the admittance $y_{i}^{\text{S}}$, which is an electrical property of the power distribution network, power and voltage constraints $s^{\text{G}}_{i, \min}$, $s^{\text{G}}_{i,\max}$, $v_{i,\min}$ and $v_{i, \max}$, which are system parameters, and cost parameters $c_{0i}$, $c_{1i}$ and $c_{2i}$. 

The branch feature vector $\bfw_{ij}$ collects all of the parameters associated with branch $(i,j)$,
\begin{equation}\label{eqn_branch_features}
    \bfw_{ij} :=
    [
    t_{ij}; y^{\text{C}}_{ij}; y^{\text{C}}_{ji}; y_{ij};
    f_{ij, \max}; f_{ji, \max};
    \theta_{ij,\min}; \theta_{ij, \max} .
    ]
\end{equation}
As in the case of the bus feature, the branch feature includes electrical circuit parameters, the transformer gain $t_{ij}$ and the admittances $y^{\text{C}}_{ij}$, $y^{\text{C}}_{ji}$ and $y_{ij}$ as well as system design parameters $f_{ij, \max}$, $f_{ji, \max}$, $\theta_{ij,\min}$ and $\theta_{ij, \max}$. An electrical grid $\bbW$ is the collection of bus features and branch features,
\begin{equation}\label{eqn_network}
    \bbW := \{\bfw_i\}_{i} \cup \{\bfw_{ij}\}_{(i,j)} .
\end{equation}
Henceforth, we adopt the following simplified notation for the OPF problem summarized in Figure \ref{fig_OPF},
\begin{alignat}{3}\label{eqn_opf_problem}
    \big[\, \bfs^\dagger(\bfr), \bfv^\dagger(\bfr) \, \big] =~ 
    &  \underset{\bfs, \bfv}{\text{argmin}}\quad    
    && C ( \bfs  ) ,                           
        \phantom{123456789012345} \\ \nonumber 
    &  \text{subject to}\quad
    && \bfg(\, \bfs,\, \bfv;\, \bfr,\, \bbW \,) ~\leq~ \mathbf{0}, 
       \\ \nonumber 
    &
    && \bfh(\, \bfs,\, \bfv;\, \bfr,\, \bbW \,) ~=~ \mathbf{0}.
\end{alignat}
In \eqref{eqn_opf_problem} the inequality constraint $\bfg(\bfs, \bfv; \bfr,{\bbW}) \leq \mathbf{0}$ represents the voltage magnitude inequalities in \eqref{eqn_voltage_limit} along with the real and imaginary parts of the constraints in \eqref{eqn_generator_limit}, \eqref{eqn_power_limit}, and \eqref{eqn_voltage_angle_difference}. The equality constraint $\bfh(\bfs, \bfv; \bfr, \bfW) = 0$ represents the real and imaginary parts of \eqref{eqn_power_balance}. In all of these constraints the power flows $\bff_{f}$ and $\bff_{t}$ are replaced by their values in \eqref{eqn_branch_from_elementwise} and \eqref{eqn_branch_to_elementwise}. Since the number of buses is $N$ and the number of branches is $M$, the number of variables in \eqref{eqn_opf_problem}, the number of equality constraints is $2N$ and the number of inequality constraints is $6N + 4M$ -- with real and imaginary parts counted as separate constraints. Power demands $\bfr$ are inputs to \eqref{eqn_opf_problem} and $\bbW$ is a parameter.


\begin{figure}[t]

{

\small

\renewcommand{\arraystretch}{1.5}

\begin{tabular}{p{0.35\linewidth} p{0.35\linewidth} p{0.15\linewidth}}
    \rowcolor[RGB]{220,220,220}
    \textbf{Variable Name} \qquad
        & \textbf{Symbol} \qquad
                    & \textbf{Location} \\
    \rowcolor[RGB]{220,220,255}
    \text{Power Demand} 
        & $r_i$ 
                    & Bus \\
    \rowcolor[RGB]{255,220,220}
    \text{Power Generation} \qquad
        & $s_i$ \qquad
                    & Bus \\
    \rowcolor[RGB]{255,220,220}
    \text{Voltage} \qquad
        & $v_{i}$ \qquad
                    & Bus \\
    \rowcolor[RGB]{220,255,220}
    \text{Flows} \qquad
        & $f_{ij}, ~ f_{ji},$ \qquad
                    & Branch \\
\end{tabular}

}

\caption{Electrical Grid Variables. Power demands $r_i$ at each bus are the input variables of the optimal power flow (OPF) problem of Figure \ref{fig_OPF} while generated powers $s_i$ and voltages $v_{i}$ are output variables. As stated in Figure \ref{fig_OPF}, the flow variables $f_{ij}$ and $f_{ji}$ at each branch are substituted for their explicit expressions in \eqref{eqn_branch_from_elementwise} and \eqref{eqn_branch_to_elementwise} to solve OPF.}

\label{fig_variables}

\end{figure}


\section{Learning with Pointwise Constraints}\label{ch4:sec:primal-dual}
Rather than solving repeatedly solving \eqref{eqn_opf_problem} every time the input variables change, we want to learn a model that can provide feasible near-optimal solutions for a \emph{distribution} $\bm\rho$ of demand vectors $\bfr$ and a given grid $\bbW$. Introduce then a learning parametrization $\Phi(\bfr; \bfA, \bbW)$ whose inputs are power demand vectors $\bfr \in \bbC^N$ and whose outputs are power generation vectors $\bfs \in \bbC^N$ along with bus voltage vectors $\bfv \in \bbC^N$,
\begin{align}\label{eqn_learning_parameterization}
    [\bfs, \bfv] 
        = \Phi(\bfr; \bfA, \bbW ).
\end{align}
The output variables are not independent, as the power flow equations can be used to determine $\bfs$ from $\bfv$. However, we prefer to soften the quality constraints during model training and perform a projection during evaluation. The electrical grid $\bbW$ [cf. \eqref{eqn_bus_features}-\eqref{eqn_network} and Figure \ref{fig_parameters}] is given in this parameterization and $\bfA$ is a set of trainable parameters -- see Section \ref{ch4:sec:gat}. 

Our goal is to find a single set of parameters $\bfA$ so that $\Phi(\bfr; \bfA, \bbW)$ provides good solutions to \eqref{eqn_opf_problem} for many different demand vectors $\bfr$ from the distribution $\bm\rho$ given a grid topology $\bbW$. To achieve this goal we replace variables $[\bfs, \bfv]$ in \eqref{eqn_opf_problem} with the output of the learning parameterization in \eqref{eqn_learning_parameterization} and further consider expectations over the demand distribution $\bm\rho$ to write the learning problem\footnote{In practice, the cost function coefficients $c_{ki}$ change daily since they are a bid made by market participants. Our parametrization does take these coefficients as inputs, but unlike $\bfr$ we assume they are constant, which is a commonly used simplification in literature.},
\begin{alignat}{3}\label{eqn_opf_learning_average}
    \bfA^\ddagger =~ 
    &  \underset{\bfA}{\text{argmin}} \quad  
    && \underset{\bfr\sim\bm\rho}{\bbE} \Big[\,
           C \Big( \Phi(\bfr; \bfA, \bbW )  \Big)  \,\Big] ,                           
       \phantom{123456789} \nonumber \\  
    &  \text{subject to} \quad
    && \underset{\bfr\sim\bm\rho}{\bbE} \Big[\,
           \bfg\Big( \Phi(\bfr; \bfA, \bbW ); \bfr, \bbW \Big)  \,\Big]
               \leq \mathbf{0}, 
       \nonumber \\ 
    &
    && \underset{\bfr\sim\bm\rho}{\bbE} \Big[\,
           \bfh \Big( \Phi(\bfr; \bfA, \bbW ); \bfr, \bbW \Big)  \,\Big]
               = \mathbf{0}.
\end{alignat}
This is an \emph{unsupervised} \emph{constrained} learning formulation. It stands in contrast to supervised learning approaches in which $\Phi(\bfr; \bfA, \bbW )$ is trained directly on solutions of \eqref{eqn_opf_problem} -- see \ref{ch4:sec:supervised} -- and it differs from standard unconstrained learning formulations in that statistical losses appears as constraints \cite{pacc, pacc_nonconvex}. 


\begin{figure}[t]

{

\small

\renewcommand{\arraystretch}{1.5}

\begin{tabular}{p{0.35\linewidth} p{0.35\linewidth} p{0.15\linewidth}}
    \rowcolor[RGB]{220,220,220}
    \textbf{Parameter Name} \qquad
        & \textbf{Symbol} \qquad
                    & \textbf{Location} \\
    \rowcolor[RGB]{255,220,220}
    \text{Bus Admittance} \qquad
        & $y_{i}^{\text{S}}$ \qquad
                    & Bus \\
    \rowcolor[RGB]{255,240,240}
    \text{Generation Bounds} \qquad
        & $s^{\text{G}}_{i, \min}, s^{\text{G}}_{i,\max}$ \qquad
                    & Bus \\
    \rowcolor[RGB]{255,240,240}
    \text{Voltage Bounds} \qquad
        & $v_{i,\min}, v_{i, \max}$ \qquad
                    & Bus \\
    \rowcolor[RGB]{255,220,220}
    \text{Cost Parameters} \qquad
        & $c_{0i}, c_{1i}, c_{2i}$ \qquad
                    & Bus \\
    \rowcolor[RGB]{220,220,255}
    \text{Line Admittance} \qquad
        & $y_{ij}$ \qquad
                    & Branch \\
    \rowcolor[RGB]{220,220,255}
    \text{Shunt Admittances} \qquad
        & $y^{\text{C}}_{ij}, ~y^{\text{C}}_{ji}$ \qquad
                    & Branch \\
    \rowcolor[RGB]{220,220,255}
    \text{Transformer Gain} \qquad
        & $t_{ij}$ \qquad
                    & Branch \\
    \rowcolor[RGB]{240,240,255}
    \text{Power Flow Bounds} \qquad
        & $f_{ij, \max}, f_{ji, \max}$ \qquad
                    & Branch \\
    \rowcolor[RGB]{240,240,255}
    \text{Phase Bounds} \qquad
        & $\theta_{ij,\min}, \theta_{ij, \max}$ \qquad
                    & Branch \\
\end{tabular}

}

\caption{Electrical Grid Parameters. An electrical grid $\bbW$ is defined by the paramters in this table associated with a list of $N$ buses $i$ and $M$ directed branches $(i,j)$. These parameters include electrical properties of buses and branches -- $y_{i}^{\text{S}}$, $y_{ij}$,  $y^{\text{C}}_{ij}$, $y^{\text{C}}_{ji}$ and $t_{ij}$ -- as well as operating range bounds -- $s^{\text{G}}_{i, \min}$, $s^{\text{G}}_{i,\max}$, $v_{i,\min}$, $v_{i, \max}$, $f_{ij, \max}$, $f_{ji, \max}$, $\theta_{ij,\min}$ and $\theta_{ij, \max}$-- and parameters of the cost function -- $c_{0i}$, $c_{1i}$ and $c_{2i}$.}

\label{fig_parameters}

\end{figure}


An alternative constrained learning formulation is to replace the \emph{expected} constraints in \eqref{eqn_opf_learning_average} with \emph{pointwise} constraints,
\begin{alignat}{3}\label{eqn_opf_learning_pointwise_not_augmented}
    \bfA^\dagger =~ 
    &  \underset{\bfA}{\text{argmin}} ~
    && \underset{\bfr\sim\bm\rho}{\bbE} \Big[\,
           C \Big(\Phi(\bfr; \bfA, \bbW )\Big)  \,\Big] ,
       \phantom{123456789} \nonumber\\
    &  \text{subject to} ~
    && \bfg\Big(\Phi(\bfr; \bfA, \bbW ); \bfr, \bbW \Big) 
               \leq \mathbf{0} 
                   ~\text{~for all~} \bfr , 
       \nonumber \\  
    &
    &&     \bfh \Big(\Phi(\bfr; \bfA, \bbW ); \bfr, \bbW \Big)  
               = \mathbf{0} 
                   ~\text{~for all~} \bfr . 
\end{alignat}
However similar, Problems \eqref{eqn_opf_learning_average} and \eqref{eqn_opf_learning_pointwise_not_augmented} are in fact disparate. In \eqref{eqn_opf_learning_average} we can violate constraints by large margins for some some realizations of $\bfr$. In the case of inequality constraints this may happen if we have some other realizations for which constraints are over-satisfied by commensurate large margins. In the case of equality constraints this can happen if constraint violations have opposite signs for different realizations of $\bfr$. In either case, the constraint is satisfied on average across realizations but violated for individual realizations. This is a serious flaw that \eqref{eqn_opf_learning_pointwise_not_augmented} overcomes by requiring constraints to hold for each individual realization. The drawback of \eqref{eqn_opf_learning_pointwise_not_augmented} is increased complexity because the number of constraints is (much) larger. It is infinite in the statistical formulation in \eqref{eqn_opf_learning_pointwise_not_augmented} and proportional to the size of the training set in the corresponding empirical formulation. Additionally, we emphasize the difference between \eqref{eqn_opf_problem} and \eqref{eqn_opf_learning_pointwise_not_augmented}; the former finds an explicit solution to a single instance of the OPF problem, while the latter finds a single set of model parameters $\bfA$ that produce OPF solutions for many different values of $\bfr$ from $\bm\rho$.

In subsequent sections we work with an augmented version of \eqref{eqn_opf_learning_pointwise} where constraints are also added as penalties to the objectve. For a precise definition consider a given demand realization vector $\bfr$ and define the quadratic penalty function
\begin{alignat}{3}\label{eqn_opf_augmentation_penalty}
    P \Big(\Phi(\bfr; \bfA, \bbW )\Big) =~ 
        &   \frac{\gamma}{2} \Big\|  \bfg_+\Big(\Phi(\bfr; \bfA, \bbW ); \bfr, \bbW \Big)  \Big\|^2 \nonumber \\
        &   ~~+ \frac{\gamma}{2}\Big\| \bfh \Big(\Phi(\bfr; \bfA, \bbW ); \bfr, \bbW\Big)  \Big\|^2 ,
\end{alignat}
where $\bfg_+(\cdot) := \max[0, \bfg(\cdot)]$ projects the output of the function $\bfg(\cdot)$ into the nonnegative reals. 

The penalty function defined in \eqref{eqn_opf_augmentation_penalty} equals zero when the pointwise constraints in \eqref{eqn_opf_learning_pointwise_not_augmented} are satisfied and it is striclty positive otherwise. It therefore follows that adding $P (\Phi(\bfr; \bfA, \bbW ))$ to the objective of \eqref{eqn_opf_learning_pointwise_not_augmented} does not alter the set of optimal arguments $\bfA^\dagger$. We have thus concluded that we can replace  \eqref{eqn_opf_learning_pointwise_not_augmented} by the equivalent formulation
\begin{alignat}{3}\label{eqn_opf_learning_pointwise}
    \bfA^\dagger =~ 
    &  \underset{\bfA}{\text{argmin}} ~
    && \underset{\bfr\sim\bm\rho}{\bbE} \Big[\,
           C \Big(\Phi(\bfr; \bfA, \bbW )\Big) + 
           P \Big(\Phi(\bfr; \bfA, \bbW )\Big)   \,\Big] ,
       \nonumber\\
    &  \text{subject to} \quad
    && \bfg\Big(\Phi(\bfr; \bfA, \bbW ); \bfr, \bbW \Big) 
               \leq \mathbf{0} 
                   ~\text{~for all~} \bfr , 
       \nonumber \\  
    &
    &&     \bfh \Big(\Phi(\bfr; \bfA, \bbW ); \bfr, \bbW \Big)  
               = \mathbf{0} 
                   ~\text{~for all~} \bfr . \hspace{-3cm}
\end{alignat}
This paper develops solutions of \eqref{eqn_opf_learning_pointwise}. We expect these solutions to learn parameters $\bfA^\dagger$ for which power and voltage estimates $[\bfs,\bfv] = \Phi(\bfr; \bfA^{\dagger}, \bbW )$ have smaller constraint violations than power and voltage estimates $[\bfs,\bfv] = \Phi(\bfr; \bfA^{\ddagger}, \bbW )$ obtained from parameters $\bfA^{\ddagger}$ learned as solutions of  \eqref{eqn_opf_learning_average}. Numerical experiments demonstrate that this is true and that learning with pointwise constraints is also better than supervised learning approaches (\ref{ch4:sec:supervised}) at yielding parameterizations that satisfy constraints for all realizations (Section \ref{ch4:sec:experiments}). 


\subsection{Dual Problem and Dual Gradient Ascent}\label{sec_duality}

Introduce dual variables $\bm\mu(\bfr) \in \bbR^{2N}$ associated with the equality constraints of \eqref{eqn_opf_learning_pointwise} and nonnegative dual variables $\bm\lambda(\bfr) \in \bbR_+^{6N + 4M}$ associated with the inequality constraints. With these definitions we write the \emph{realization} Lagrangian as
\begin{alignat}{3}\label{eqn_realization_lagrangian}
     l \Big( \bfA, \bm\lambda(\bfr), \bm\mu(\bfr), \bfr \Big) 
           & ~=~ C \Big( \Phi(\bfr; \bfA, \bbW )  \Big) 
                + P \Big(\Phi(\bfr; \bfA, \bbW )\Big)
                      \nonumber \\
           & ~+~ \bm\lambda(\bfr) \times
                     \bfg \Big( \Phi ( \bfr; \bfA, \bbW ); \bfr, \bbW \Big)
                         \nonumber \\
           & ~+~ \bm\mu(\bfr) \times
                     \bfh \Big( \Phi ( \bfr; \bfA, \bbW ); \bfr, \bbW \Big) ,
\end{alignat}
whose expectation over the demand distribution $\bm\rho$ yields the Lagrangian of the optimization problem in \eqref{eqn_opf_learning_pointwise},
\begin{alignat}{3}\label{eqn_opf_lagrangian}
    L(\bfA, \bm\lambda, \bm\mu)  
    & ~=~\underset{\bfr\sim\bm\rho}{\bbE} \Big[\, l\Big(\bfA, \bm\lambda(\bfr), \bm\mu(\bfr), \bfr \Big) \,\Big]  
       .  
\end{alignat}
The Lagrangian and realization Lagrangian are also functions of the electrical network $\bbW$ but we omit this dependence to simplify notation.

From the Lagrangian in \eqref{eqn_opf_lagrangian} we define the dual function $d(\bm\lambda, \bm\mu)$ as the minimum value of the Lagrangian over parameters $\bfA$. We write the dual function as $d(\bm\lambda, \bm\mu) = L(\bfA(\bm\lambda, \bm\mu), \bm\lambda, \bm\mu)$ where the parameters $\bfA(\bm\lambda, \bm\mu)$ are minimizers of the Lagrangian for a given pair of multipliers $(\bm\lambda, \bm\mu)$,
\begin{align}\label{eqn_lagrangian_minimizers}
    \bfA(\bm\lambda, \bm\mu)
        ~ = ~ \underset{\bfA}{\text{argmin}} ~
                  L(\bfA, \bm\lambda, \bm\mu) .
\end{align}
Further define the dual problem as the dual function maximization. That is, search for dual variables $(\bm\lambda^\dagger,\bm\mu^\dagger)$ such that,
\begin{align}\label{eqn_optimal_dual_variables}
    (\bm\lambda^\dagger,\bm\mu^\dagger)
         ~=~ \underset{\bm\lambda \geq \mathbf{0}, ~\bm\mu}{\text{argmax}} ~
                  d(\bm\lambda, \bm\mu) 
         ~=~ \underset{\bm\lambda \geq \mathbf{0}, ~\bm\mu}{\text{argmax}} ~
                  L(\bfA(\bm\lambda, \bm\mu), \bm\lambda, \bm\mu) .
\end{align}
The determination of the Lagrangian minimizers $\bfA(\bm\lambda, \bm\mu)$ in \eqref{eqn_lagrangian_minimizers} is a standard unconstrained learning problem. The realization Lagrangian in \eqref{eqn_realization_lagrangian} is the loss associated with realization $\bfr$ and the Lagrangian in \eqref{eqn_opf_lagrangian} is the average loss. We can thus solve the (unconstrained) learning problem in \eqref{eqn_lagrangian_minimizers} in lieu of the constrained learning problem in \eqref{eqn_opf_learning_pointwise}. Since different multipliers  $[\bm\lambda,\bm\mu]$ yield different parameters $\bfA(\bm\lambda, \bm\mu)$, their judicious choice is necessary. Duality theory indicates that optimal multipliers $(\bm\lambda^\dagger,\bm\mu^\dagger)$ are the best choice \cite[Ch. 5]{boyd2004convex}. We therefore propose here to learn parameters $\bfA(\bm\lambda^\dagger, \bm\mu^\dagger)$ obtained by minimizing the Lagrangian associated with optimal multipliers $(\bm\lambda^\dagger,\bm\mu^\dagger)$, 
\begin{align}\label{eqn_lagrangian_minimizers_optimal}
    \bfA(\bm\lambda^\dagger, \bm\mu^\dagger)
        ~ = ~ \underset{\bfA}{\text{argmin}} ~
                  L(\bfA, \bm\lambda^\dagger, \bm\mu^\dagger) .
\end{align}
Since the inequality constraint function $\bfg$ is not convex and the equality constraint function $\bfh$ is not linear, strong duality \cite[Ch. 5]{boyd2004convex} does not necessarily hold for \eqref{eqn_opf_learning_pointwise}. This means we cannot claim the equivalence of $\bfA(\bm\lambda^\dagger, \bm\mu^\dagger)$ and $\bfA^\dagger$. We say that \eqref{eqn_lagrangian_minimizers_optimal} is the Lagrangian relaxation of \eqref{eqn_opf_learning_pointwise} -- see also Remark \ref{rmk_strong_duality}. 

To find $\bfA(\bm\lambda^\dagger, \bm\mu^\dagger)$ we need to find the pair of optimal multipliers $(\bm\lambda^\dagger, \bm\mu^\dagger)$ and solve \eqref{eqn_lagrangian_minimizers_optimal}. The latter is a standard unconstrained learning problem -- a particular case of \eqref{eqn_lagrangian_minimizers}, in fact. To find optimal multipliers $(\bm\lambda^\dagger, \bm\mu^\dagger)$ we recall that supergradients of the dual function are given by the  constraint slacks associated with corresponding Lagrangian minimizers \cite[Ch. 5]{boyd2004convex} ,
\begin{align}\label{eqn_dual_function_supergradients}
    \frac{\partial}{\partial\bm\lambda(\bfr)}  d(\bm\lambda, \bm\mu)
        ~=~ \bfg\Big( \Phi(\bfr; \bfA(\bm\lambda, \bm\mu), \bbW ); \bfr, \bbW \Big),
    \nonumber \\
    \frac{\partial }{\partial\bm\mu(\bfr)}  d(\bm\lambda, \bm\mu)
        ~=~ \bfh\Big( \Phi(\bfr; \bfA(\bm\lambda, \bm\mu), \bbW ); \bfr, \bbW \Big),
\end{align}
We can now follow these supergradients to find the optimal multipliers $(\bm\lambda^\dagger, \bm\mu^\dagger)$. Formally, introduce an iteration index $t$, a step-size $\alpha$ and proceed to update dual variables according to the recursion,
\begin{alignat}{5}\label{eqn_supergradient_ascent}
    &\bm\lambda_{t+1}(\bfr) = \Big [ \,  
          && \bm\lambda_{t}(\bfr) 
                &&+ \alpha 
                        \bfg &&\Big( \Phi(\bfr; \bfA(\bm\lambda_{t}, \bm\mu_{t}), \bbW ); \bfr, \bbW \Big)
                                 \,\Big]_+ ,
    \nonumber \\
    &\bm\mu_{t+1}(\bfr) = 
          && \bm\mu_{t}(\bfr) 
                &&+ \alpha 
                        \bfh &&\Big( \Phi(\bfr; \bfA(\bm\lambda_{t}, \bm\mu_{t}), \bbW ); \bfr, \bbW \Big)
                             .
\end{alignat}
Because the constraint slacks associated with the Lagrangian minimizers $\bfA(\bm\lambda_{t}, \bm\mu_{t})$ are supergradients of the dual function at $(\bm\lambda_{t}, \bm\mu_{t})$ [cf. \eqref{eqn_dual_function_supergradients}], the recursion in \eqref{eqn_supergradient_ascent} defines a supergradient ascent algorithm for the dual function $d(\bm\lambda, \bm\mu)$. Since the dual function is concave on the dual variables $(\bm\lambda, \bm\mu)$ \cite[Ch. 5]{boyd2004convex}, \eqref{eqn_supergradient_ascent} is guaranteed to find optimal multipliers. Observe that in \eqref{eqn_supergradient_ascent}, the operator $[\bm\lambda]_+ = \max(\bm 0, \bm\lambda)$ projects multiplier iterates $\bm\lambda_{t+1}(\bfr)$ to the nonnegative orthant. This is needed because we have defined the multipliers $\bm\lambda(\bfr) \in \bbR_+^{6N + 4M}$ associated with inequality constraints to be nonnegative.


\begin{figure}

\centering

\addtolength{\jot}{0.5em}

\hrule

{

\small

\begin{alignat}{3}
    &\textbf{(P)} \quad 
        && \text{Determine parameter $\bfA$ that minimizes the Lagrangian}
                               \nonumber \\
    & 
        && \bfA_{t+1} 
               ~=~ \bfA(\bm\lambda_{t}, \bm\mu_{t}) 
               ~=~ \underset{\bfA}{\text{argmin}} ~
                       L(\bfA, \bm\lambda_{t}, \bm\mu_{t})  
                           \tag{\ref{eqn_lagrangian_minimizers}}
                               \nonumber \\
    &\textbf{(D)} \quad 
        && \text{Update dual variables for all realizations of demand vector $\bfr$}
                               \nonumber \\
    & \quad
        && \bm\lambda_{t+1}(\bfr) 
               = \Big [ \, \bm\lambda_{t}(\bfr) 
                     ~+~ \alpha 
                             \bfg\Big(\Phi(\bfr; \bfA_{t+1}, \bbW ); \bfr, \bbW  \Big)
                                 \,\Big]_+ 
                                       \nonumber \\
    &
        && \bm\mu_{t+1}(\bfr) 
               = \phantom{\Big [ \, } 
                       \bm\mu_{t}(\bfr) \hspace{-1pt}
                          ~+~ \alpha 
                                   \bfh\Big(\Phi(\bfr; \bfA_{t+1}, \bbW ); \bfr, \bbW \Big)
                                       \phantom{\, \Big ]} 
                                       \tag{\ref{eqn_supergradient_ascent}}
                                           \nonumber 
\end{alignat}    

}

\hrule
    
\caption{Dual Gradient Ascent Iteration. The primal step (P) determines parameters $\bfA_{t+1} = \bfA(\bm\lambda_{t}, \bm\mu_{t})$ that minimize the Lagrangian $L(\bfA, \bm\lambda_{t}, \bm\mu_{t})$ [cf. \eqref{eqn_lagrangian_minimizers}]. The dual step (D) updates dual variables by following constraint slacks [cf. \eqref{eqn_supergradient_ascent}]. When we exit dual gradient ascent after $T$ iterations we adopt $\bfA_{T}$ as a proxy for the optimal argument $\bfA^\dagger$ [cf. \eqref{eqn_opf_learning_pointwise}].}

\label{fig_dual_gradient_ascent}
        
\end{figure}


An iteration of the dual gradient ascent algorithm is summarized in Figure \ref{fig_dual_gradient_ascent}. In the primal step (P) we determine parameters $\bfA_{t+1} = \bfA(\bm\lambda_{t}, \bm\mu_{t})$ that minimize the Lagrangian $L(\bfA, \bm\lambda_{t}, \bm\mu_{t})$ for given multipliers $(\bm\lambda_{t}, \bm\mu_{t})$. This is an ancillary computation that permits updating of dual variables in the dual step (D). When we exit dual gradient ascent after $T$ iterations the multipliers $(\bm\lambda_{T}, \bm\mu_{T})$ are approximations of the optimal multipliers $(\bm\lambda^\dagger,\bm\mu^\dagger)$ and the primal iterate $\bfA_{T}$ is an approximation of the Lagrangian minimizer $\bfA(\bm\lambda^\dagger, \bm\mu^\dagger)$ which we have accepted as a proxy for the optimal argument $\bfA^\dagger$. 


\begin{remark} \label{rmk_strong_duality}\normalfont

The Lagrangian of \eqref{eqn_opf_learning_pointwise} is akin to the augmented Lagrangian of \eqref{eqn_opf_learning_pointwise_not_augmented}. This is important here because under mild regularity conditions augmented Lagrangians of nonconvex problems recover optimal primal variables \cite{Wierzbicki_Kurcyusz}. For the OPF problem we consider here, we can claim $\bfA(\bm\lambda^\dagger, \bm\mu^\dagger) \equiv \bfA^\dagger$ provided that the penalty coefficient $\gamma$ is sufficiently large. I.e., operating in the dual domain carries no loss of optimality as we can recover optimal primal variables $\bfA^\dagger$ from Lagrangian minimizers $\bfA(\bm\lambda^\dagger, \bm\mu^\dagger)$. This is in contrast to working with the (not augmented) Lagrangian of \eqref{eqn_opf_learning_pointwise_not_augmented} for which we cannot make strong duality claims. This is the main motivation for working with the penalty formulation \eqref{eqn_opf_learning_pointwise} instead of the initial formulation \eqref{eqn_opf_learning_pointwise_not_augmented}.

\end{remark}


\subsection{Training in the Dual Domain}\label{sec_dual_training}

To derive a training algorithm we modify dual gradient descent algorithm (Figure \ref{fig_dual_gradient_ascent}) by: (i) Considering stochastic approximations of the Lagrangian $L(\bfA, \bm\lambda, \bm\mu)$. (ii) Replacing the minimization in the primal step (P) with gradient descent iterations on the stochastic Lagrangian. (iii) Updating a random subset of dual variables in the dual step (D).

Consider then a batch $B_t$ with $B$ samples $\bfr\sim\bm\rho$ drawn from the data distribution at iteration $t$. Based on the entries in this batch we define the stochastic Lagrangian $L_{B_t}(\bfA, \bm\lambda, \bm\mu)$ as the sample estimate
\begin{alignat}{3}\label{eqn_opf_lagrangian_empirical}
    L_{B_t}(\bfA, \bm\lambda, \bm\mu) ~=~
    & \frac{1}{B}\sum_{\bfr \in B_t}
          l \Big( \bfA, \bm\lambda(\bfr), \bm\mu(\bfr), \bfr \Big).
\end{alignat}
We now introduce a step size $\beta$ and proceed to update variables $\bfA_t$ by following the opposite of the gradient of the Lagrangian with respect to primal parameters,
\begin{align}\label{eqn_lagrangian_gradient_descent}
    \bfA_{t+1}~=~ 
        \bfA_{t} - \beta \frac{\partial}{\partial \bfA} L_{B_t}(\bfA_{t}, \bm\lambda_{t}, \bm\mu_{t}) .
\end{align}
We now consider a possibly different batch $C_t$ of demand vector realizations $\bfr\sim\bf\rho$ and proceed to update the dual variables $\bm\lambda_{t}(\bfr)$ and $\bm\mu_{t+1}(\bfr)$ associated with realizations $\bfr \in C_t$. We do so by following the constraint slacks associated with parameter $\bfA_{t+1}$,
\begin{alignat}{5}\label{eqn_supergradient_ascent_stochastic}
    &\bm\lambda_{t+1}(\bfr) = \Big [ 
          && \bm\lambda_{t}(\bfr) 
                &&+ \alpha 
                        \bfg &&\Big( \Phi(\bfr; \bfA_{t+1}, \bbW ); \bfr, \bbW \Big)
                                 \Big]_+ ,
    \nonumber \\
    &\bm\mu_{t+1}(\bfr) = 
          && \bm\mu_{t}(\bfr) 
                &&+ \alpha 
                        \bfh &&\Big( \Phi(\bfr; \bfA_{t+1}, \bbW ); \bfr, \bbW \Big)
      .
\end{alignat}
The update in \eqref{eqn_supergradient_ascent_stochastic} differs from \eqref{eqn_supergradient_ascent} in that we update dual variables associated with demand vectors in the batch $C_t$ instead of updating dual variables corresponding to all realizations of $\bfr$. These updates are also different in that \eqref{eqn_supergradient_ascent_stochastic} evaluates constraint slacks associated with parameters $\bfA_{t+1}$ generated by the gradient descent step in \eqref{eqn_lagrangian_gradient_descent} instead of the Lagrangian minimizers $\bfA(\bm\lambda_{t}, \bm\mu_{t})$ used in \eqref{eqn_supergradient_ascent}. If the variables $\bfA_{t+1}$ are close to Lagrangian minimizers $\bfA(\bm\lambda_{t}, \bm\mu_{t})$, \eqref{eqn_supergradient_ascent_stochastic} is close to a stochastic block coordinate ascent algorithm on the \emph{concave} dual function. We thus expect convergence of \eqref{eqn_supergradient_ascent_stochastic} because  block coordinate ascent algorithms converge for concave functions \cite{Mokhtari_Koppel}.


\begin{figure}

\centering

\addtolength{\jot}{0.5em}

\hrule

{

\small

\begin{alignat}{3}
    &\textbf{(P)} \quad 
        && \text{Draw primal batch $B_t$ and descend on stochastic Lagrangian} 
                               \nonumber \\
    & 
        && \bfA_{t+1}~=~ 
               \bfA_{t} 
                   - \beta \frac{\partial}{\partial \bfA} 
                         L_{B_t}(\bfA_{t}, \bm\lambda_{t}, \bm\mu_{t}) 
                             \tag{\ref{eqn_lagrangian_gradient_descent}}
                                 \nonumber \\
    &\textbf{(D)} \quad 
        && \text{Update dual variables for demand vectors $\bfr$ in dual batch $C_t$}
                               \nonumber \\
    & \quad
        && \bm\lambda_{t+1}(\bfr) 
               = \Big [ \, \bm\lambda_{t}(\bfr) 
                     ~+~ \alpha 
                             \bfg\Big(\Phi(\bfr; \bfA_{t+1}, \bbW ); \bfr, \bbW  \Big)
                                 \,\Big]_+ 
                                       \nonumber \\
    &
        && \bm\mu_{t+1}(\bfr) 
               = \phantom{\Big [ \, } 
                       \bm\mu_{t}(\bfr) \hspace{-1pt}
                          ~+~ \alpha 
                                   \bfh\Big(\Phi(\bfr; \bfA_{t+1}, \bbW ); \bfr, \bbW \Big)
                                       \phantom{\, \Big ]} 
                                       \tag{\ref{eqn_supergradient_ascent_stochastic}}
                                           \nonumber 
\end{alignat}    

}

\hrule
    
\caption{Training Iteration. Samples in the primal batch $B_t$ define the primal stochastic gradient descent step [cf. \eqref{eqn_opf_lagrangian_empirical} and \eqref{eqn_lagrangian_gradient_descent}]. The dual step (D) follows constraint slacks associated with the updated primal variable $\bfA_{t+1}$ to update dual variables associated with demand realizations in the dual batch $C_t$ [cf. \eqref{eqn_supergradient_ascent_stochastic}]. After $T$ iterations we accept $\bfA_T$ as an approximation of $\bfA^\dagger$ [cf. \eqref{eqn_opf_learning_pointwise}].} 

\label{fig_training_iteration}
        
\end{figure}


An iteration of the resulting training algorithm is summarized in Figure \ref{fig_training_iteration}. In the primal step we select a (primal) batch of realizations $B_t$ which we use to define the stochastic Lagrangian $L_{B_t}(\bfA, \bm\lambda, \bm\mu)$ [cf. \eqref{eqn_opf_lagrangian_empirical}]. We then update the primal parameter $\bfA_t$ by descending along the primal gradient of the stochastic Lagrangian [cf. \eqref{eqn_lagrangian_gradient_descent}]. This is equivalent to a primal stochastic gradient descent step taken on the Lagrangian $L(\bfA, \bm\lambda, \bm\mu)$ [cf. \eqref{eqn_opf_lagrangian}]. In the dual step we select a possibly different (dual) batch of samples $C_t$ and update dual variables associated with demand realizations $\bfr \in C_t$. These updates follow the constraint slacks associated with the updated primal variable $\bfA_{t+1}$ [cf. \eqref{eqn_supergradient_ascent_stochastic}]. When we exit the training algorithm after $T$ iterations we accept $\bfA_T$ as an approximation of the optimal parameter $\bfA^\dagger$.


\subsection{Dual Training with Shared Multipliers}\label{sec_shared_multipliers}

It is instructive to better to formulate a version of \eqref{eqn_opf_learning_pointwise} with expected constraints
\begin{alignat}{3}\label{eqn_opf_learning_average_with_penalty}
    \bfA^{\ddagger\ddagger} = 
    &  \underset{\bfA}{\text{argmin}} \quad  
    && \underset{\bfr\sim\bm\rho}{\bbE} \Big[\,
           C \Big(\Phi(\bfr; \bfA, \bbW )\Big) + 
           P \Big(\Phi(\bfr; \bfA, \bbW )\Big)   \,\Big] ,
       \nonumber \\  
    &  \text{subject to} \quad
    && \underset{\bfr\sim\bm\rho}{\bbE} \Big[\,
           \bfg\Big( \Phi(\bfr; \bfA, \bbW ); \bfr, \bbW \Big)  \,\Big]
               \leq \mathbf{0}, 
       \nonumber \\ 
    &
    && \underset{\bfr\sim\bm\rho}{\bbE} \Big[\,
           \bfh \Big( \Phi(\bfr; \bfA, \bbW ); \bfr, \bbW \Big)  \,\Big]
               = \mathbf{0}.
\end{alignat}
This problem is \emph{not} equivalent to \eqref{eqn_opf_learning_average} because the penalty function $P (\Phi(\bfr; \bfA, \bbW ))$ is pointwise but the constraints are required in expectation. We think of  \eqref{eqn_opf_learning_average_with_penalty} as a lower complexity alternative to \eqref{eqn_opf_learning_pointwise} that mitigates the drawbacks of \eqref{eqn_opf_learning_average}. Requiring constraints on average reduces the number of dual variables whereas the penalty term encourages -- as opposed to requiring -- satisfaction of constraints for each individual realization of the demand vector $\bfr$. 


\begin{figure}

\centering

\addtolength{\jot}{0.5em}

\hrule

{

\small

\begin{alignat}{3}
    &\textbf{(P)} \quad 
        && \text{Draw primal batch $B_t$ and descend on stochastic Lagrangian} 
                               \nonumber \\
    & 
        && \bfA_{t+1}~=~ 
    \bfA_{t+1}~=~ 
        \bfA_{t} - \beta \frac{\partial}{\partial \bfA} 
             \bar L_{B_t}(\bfA, \bar{\bm\lambda}_t, \bar{\bm\mu}_t) .
                             \tag{\ref{eqn_lagrangian_gradient_descent_shared}}
                                 \nonumber \\
    &\textbf{(D)} \quad 
        && \text{Draw dual batch $C_t$ and ascend on stochastic dual gradient}
                               \nonumber \\
    & \quad
        && \bar{\bm\lambda}_{t+1}
               = \bigg [ \, \bar{\bm\lambda}_{t}
                     ~+~ \frac{\alpha }{C} \sum_{\bfr \in C_t} 
                             \bfg\Big(\Phi(\bfr; \bfA_{t+1}, \bbW ); \bfr, \bbW  \Big)
                                 \,\bigg]_+ 
                                       \nonumber \\
    &
        && \bar{\bm\mu}_{t+1}
               = \phantom{\Big [ \, } 
                       \bar{\bm\mu}_{t} \hspace{-1pt}
                          ~+~ \frac{\alpha }{C} \sum_{\bfr \in C_t} 
                                   \bfh\Big(\Phi(\bfr; \bfA_{t+1}, \bbW ); \bfr, \bbW \Big)
                                       \phantom{\, \Big ]}  
                                       \tag{\ref{eqn_supergradient_ascent_stochastic_shared}}
                                           \nonumber 
\end{alignat}    

}

\hrule
    
\caption{Training Iteration for Shared Multipliers. In contrast to Figure \ref{fig_training_iteration} we use shared multipliers $\bar{\bm\lambda}_t$ and $\bar{\bm\mu}_t$ to build a stochastic Lagrangian in the primal step (P) [cf. \eqref{eqn_opf_lagrangian_empirical_shared} and \eqref{eqn_lagrangian_gradient_descent_shared}]. In the dual step (D) we update the shared multipliers by ascending on stochastic gradients of the dual function [cf. \eqref{eqn_supergradient_ascent_stochastic_shared}]. After $T$ iterations the parameter $\bfA_T$ is an approximation of $\bfA^{\ddagger\ddagger}$ [cf. \eqref{eqn_opf_learning_average_with_penalty}].} 

\label{fig_training_iteration_shared}
        
\end{figure}


To write down the Lagrangian of \eqref{eqn_opf_learning_average_with_penalty} we make use of the definition of the realization Lagrangian in \eqref{eqn_realization_lagrangian}. Using this definition and introducing the dual variable $\bar{\bm\mu}\in \bbR^{2N}$ associated with the equality constraint of \eqref{eqn_opf_learning_average_with_penalty} and the nonnegative dual variable $\bar{\bm\lambda}\in \bbR_+^{6N + 4M}$ associated with the inequality constraint we can write
\begin{alignat}{3}\label{eqn_opf_lagrangian_shared}
    \bar L(\bfA, \bar{\bm\lambda}, \bar{\bm\mu})  
    & ~=~\underset{\bfr\sim\bm\rho}{\bbE} \Big[\, l\Big(\bfA, \bar{\bm\lambda}, \bar{\bm\mu}, \bfr \Big) \,\Big] 
       .  
\end{alignat}
According to this definition, $\bar L(\bfA, \bar{\bm\lambda}, \bar{\bm\mu})$  is a particular case of $L(\bfA, \bm\lambda, \bm\mu)$ in which we require $\bm\lambda(\bfr) = \bar{\bm\lambda}$ and $\bm\mu(\bfr) = \bar{\bm\mu}$ for all $\bfr$. I.e., it is the same Lagrangian with the restriction that the multipliers $\bm\lambda(\bfr) = \bar{\bm\lambda}$ and $\bm\mu(\bfr) = \bar{\bm\mu}$ are \emph{shared} by all realizations of $\bfr$.

To train with shared multipliers we consider a version of the stochastic Lagrangian with shared multipliers. Thus, for a a batch $B_t$ of demand vector realizations we define 
\begin{alignat}{3}\label{eqn_opf_lagrangian_empirical_shared}
    \bar L_{B_t}(\bfA, \bar{\bm\lambda}, \bar{\bm\mu}) ~=~
    & \frac{1}{B}\sum_{\bfr \in B_t}
          l \Big( \bfA, \bar{\bm\lambda}, \bar{\bm\mu}, \bfr \Big),
\end{alignat}
and replace $\bar L_{B_t}(\bfA, \bar{\bm\lambda}_t, \bar{\bm\mu}_t)$ for $L_{B_t}(\bfA, \bm\lambda_t, \bm\mu_t)$ in the primal iteration in \eqref{eqn_lagrangian_gradient_descent},
\begin{align}\label{eqn_lagrangian_gradient_descent_shared}
    \bfA_{t+1}~=~ 
        \bfA_{t} - \beta \frac{\partial}{\partial \bfA} 
             \bar L_{B_t}(\bfA, \bar{\bm\lambda}_t, \bar{\bm\mu}_t) .
\end{align}
For the dual iteration recall that $\bar{\bm\lambda}$ and $\bar{\bm\mu}$ are associated with \emph{expectation} constraints. Thus, stochastic gradient ascent steps for the dual function are obtained by evaluating \emph{average} constraint slacks across a batch $C_t$ of demand vector realizations. Thus, if we let $C$ denote the number of samples in the dual batch we update dual variables as,
\begin{alignat}{5}\label{eqn_supergradient_ascent_stochastic_shared}
    &\bar{\bm\lambda}_{t+1}= \bigg [ 
          && \bar{\bm\lambda}_{t}
                && + \frac{\alpha}{C} \sum_{\bfr \in C_t} 
                            \bfg && \Big( \Phi(\bfr; \bfA_{t+1}, \bbW ); \bfr, \bbW \Big)
                                 \bigg]_+ ,
    \nonumber \\
    &\bar{\bm\mu}_{t+1} = 
          && \bar{\bm\mu}_{t}
                && + \frac{\alpha }{C} \sum_{\bfr \in C_t} 
                            \bfh && \Big( \Phi(\bfr; \bfA_{t+1}, \bbW ); \bfr, \bbW \Big)      .
\end{alignat}
In this stochastic gradient ascent algorithm the batch $C_t$ is used to estimate the gradient of the dual function of problem \eqref{eqn_opf_learning_average_with_penalty}. This is different from the role of the batch $C_t$ in the stochastic block coordinate ascent algorithm in \eqref{eqn_supergradient_ascent} where updates follow the actual gradients of the dual function of \eqref{eqn_opf_learning_pointwise} over a random subset of multipliers as determined by the batch $C_t$.

An iteration of the algorithm for training with shared multipliers is shown in Figure \ref{fig_training_iteration_shared}. The primal step (P) is analogous to the primal step in Figure \ref{fig_training_iteration} which corresponds to the case of pointwise multipliers. The difference is that we use shared multipliers $\bar{\bm\lambda}_t$ and $\bar{\bm\mu}_t)$ to build a (shared) stochastic Lagrangian instead of pointwise multipliers $\bm\lambda_{t}$ and $\bm\mu_{t}$ to build a (pointwise) stochastic Lagrangian. In the dual step (D) we update the shared multipliers by ascending on stochastic gradients of the dual function. After $T$ iterations the parameter $\bfA_T$ is an approximation of $\bfA^{\ddagger\ddagger}$ [cf. \eqref{eqn_opf_learning_average_with_penalty}].


\begin{figure}

\centering

\addtolength{\jot}{0.5em}

\hrule

{

\small

\begin{alignat}{3}
    & \textbf{(I)} \quad 
        && \text{Compute pointwise dual variables for $\bfr \in B_t$ and $\bfr \in C_t$} 
               \nonumber \\
    & 
        && \bm\lambda_{t} (\bfr)= \bar{\bm\lambda}_{t} + \Delta\bm\lambda_{t}(\bfr)
               \nonumber \\
    & 
        && \bm\mu_{t}     (\bfr)= \bar{\bm\mu}_{t}     + \Delta\bm\mu_{t}(\bfr) 
               \tag{\ref{eqn_hybrid_multipliers}}
               \nonumber \\
    &\textbf{(P)} \quad 
        && \text{Draw primal batch $B_t$ and descend on stochastic Lagrangian} 
                               \nonumber \\
    & 
        && \bfA_{t+1}~=~ 
               \bfA_{t} 
                   - \beta \frac{\partial}{\partial \bfA} 
                         L_{B_t}(\bfA_{t}, \bm\lambda_{t}, \bm\mu_{t}) 
                             \tag{\ref{eqn_lagrangian_gradient_descent}}
                                 \nonumber \\
    &\textbf{(D$_{\textbf{P}}$)} \quad 
        && \text{Update pointwise duals for demands $\bfr$ in dual batch $C_t$}
                               \nonumber \\
    & \quad
        && \bm\lambda_{t+1}(\bfr) 
               = \Big [ \, \bm\lambda_{t}(\bfr) 
                     ~+~ \alpha 
                             \bfg\Big(\Phi(\bfr; \bfA_{t+1}, \bbW ); \bfr, \bbW  \Big)
                                 \,\Big]_+ 
                                       \nonumber \\
    &
        && \bm\mu_{t+1}(\bfr) 
               = \phantom{\Big [ \, } 
                       \bm\mu_{t}(\bfr) \hspace{-1pt}
                          ~+~ \alpha 
                                   \bfh\Big(\Phi(\bfr; \bfA_{t+1}, \bbW ); \bfr, \bbW \Big)
                                       \phantom{\, \Big ]} 
                                       \tag{\ref{eqn_supergradient_ascent_stochastic}}
                                           \nonumber \\
    &\textbf{(D$_{\textbf{S}}$)} \quad 
        && \text{Update shared duals by asceding on stochastic dual gradient}
                               \nonumber \\
    & \quad
        && \bar{\bm\lambda}_{t+1}
               = \bigg [ \, \bar{\bm\lambda}_{t}
                     ~+~ \frac{\alpha }{C} \sum_{\bfr \in C_t} 
                             \bfg\Big(\Phi(\bfr; \bfA_{t+1}, \bbW ); \bfr, \bbW  \Big)
                                 \,\bigg]_+ 
                                       \nonumber \\
    &
        && \bar{\bm\mu}_{t+1}
               = \phantom{\Big [ \, } 
                       \bar{\bm\mu}_{t} \hspace{-1pt}
                          ~+~ \frac{\alpha }{C} \sum_{\bfr \in C_t} 
                                   \bfh\Big(\Phi(\bfr; \bfA_{t+1}, \bbW ); \bfr, \bbW \Big)
                                       \phantom{\, \Big ]}  
                                       \tag{\ref{eqn_supergradient_ascent_stochastic_shared}}
                                           \nonumber \\
    & \textbf{(O)} \quad 
        && \text{Update dual variable deviations for over dual batch $C_t$} 
              \nonumber \\
    & 
        && \Delta\bm\lambda_{t+1}(\bfr) ~=~ \bm\lambda_{t+1}(\bfr) ~-~ \bar{\bm\lambda}_{t+1}
               \nonumber \\
    & 
        && \Delta\bm\mu_{t+1}(\bfr)     ~=~ \bm\mu_{t+1} (\bfr)    ~-~ \bar{\bm\mu}_{t+1}
               \tag{\ref{eqn_hybrid_multipliers}}
               \nonumber 
\end{alignat}    

}

\hrule
    
\caption{Training Iteration for Hybrid Multipliers. The input step (I) evaluates pointwise multipliers as sums of shared multipliers and multiplier deviations [cf. \eqref{eqn_hybrid_multipliers}]. The primal step (P) updates $\bfA_t$ with a stochastic gradient descent iteration  [cf. \eqref{eqn_lagrangian_gradient_descent}]. The dual steps (D$_{\text{P}}$) and (D$_{\text{S}}$) update pointwise [cf. \eqref{eqn_supergradient_ascent_stochastic}] and shared [cf. \eqref{eqn_supergradient_ascent_stochastic_shared}] multipliers, respectively. Step (O) computes dual variable deviations [cf. \eqref{eqn_hybrid_multipliers}]. After $T$ iterations, $\bfA_T$ approximates $\bfA^{\dagger}$ [cf. \eqref{eqn_opf_learning_average_with_penalty}].} 

\label{fig_training_iteration_hybrid}
        
\end{figure}


\begin{figure*}[t]

\centering

\small

\renewcommand{\arraystretch}{1.3}

\setlength{\tabcolsep}{3.34mm}

\begin{tabular}
{ 
    l 
    r 
    >{\raggedleft\arraybackslash}p{0.8cm}
    >{\raggedleft\arraybackslash}p{0.8cm}
    r 
    r
    >{\raggedleft\arraybackslash}p{0.8cm}
    >{\raggedleft\arraybackslash}p{0.8cm}
    r 
    r 
    >{\raggedleft\arraybackslash}p{0.8cm} 
    >{\raggedleft\arraybackslash}p{1.1cm}
    >{\raggedleft\arraybackslash}p{1.1cm}
    >{\raggedleft\arraybackslash}p{1.1cm}  
}

\midrule

& \multicolumn{4}{c}{Optimality Gap (\%)} 
& \multicolumn{4}{c}{Mean Violation (\%)} 
& \multicolumn{5}{c}{Max Violation  (\%)} \\

\cmidrule(lr){3-4}\cmidrule(lr){7-8}\cmidrule(lr){11-14}

&& Mean & Std &&& Mean & Std &&& Mean & Std & P95 & Max \\

\midrule

Dual-P   && 1.07 & 1.69 &&& 0.69 &  1.01 &&&   5.68 &     8.84 &    17.05 &   247.38 \\
Dual-H   && 1.38 & 1.99 &&& 0.90 & 0.68  &&&  10.82 &    15.58 &    39.25 &   169.00 \\
Dual-S   && 1.48 & 2.16 &&& 9.82 & 21.56 &&& 430.47 & 1,164.93 & 2,676.58 & 9,575.20 \\
MSE      && 0.22 & 0.66 &&& 2.40 &  3.09 &&&  18.97 &    40.57 &    64.69 &   858.42 \\
MSE+Pen. && 0.77 & 1.16 &&& 8.30 & 15.17 &&& 193.33 &   388.05 & 1,098.28 & 1,872.07 \\

\midrule

\end{tabular}

\caption{Optimality and Constraint Violation Aggregate Distribution. The optimality gap, mean violation, and maximum violation are computed for each sample in the test sets and then aggregated across all samples. We report several metrics, including the mean, standard deviation, maximum, and 95th percentile. The statistics are computed across all samples in all power systems. See Section~\ref{sec_evaluation} for more details.}

\label{ch4:tab:model_summary}

\end{figure*}


\subsection{Dual Training with Hybrid Multipliers}\label{sec_hybrid_multipliers}

Training with pointwise multipliers (Section \ref{sec_dual_training}) enforces constraints for all realizations at the cost of a high dimensional dual function. Training with shared multipliers (Section \ref{sec_shared_multipliers}) results in a low dimensional dual function at the cost of encouraging but not enforcing constraint satisfaction for all realizations. As an intermediate compromise we consider a hybrid algorithm mixing multipliers associated with average and pointwise constraints. We do that by writing pointwise multipliers as sums of shared mutlipliers and corresponding deviations 
\begin{align}\label{eqn_hybrid_multipliers}
    \bm\lambda_{t} (\bfr) = \bar{\bm\lambda}_{t} + \Delta\bm\lambda_{t}(\bfr), \qquad
           \bm\mu_{t}(\bfr) = \bar{\bm\mu}_{t}     + \Delta\bm\mu_{t}(\bfr) 
\end{align}
To train with hybrid multipliers we update primal variables using pointwise multipliers as in Figure \ref{fig_training_iteration}, pointwise multipliers $\bm\lambda_{t} (\bfr)$ and $\bm\mu_{t}(\bfr)$ are updated with block coordinate descent iterations also as in Figure \ref{fig_training_iteration}, and shared multipliers $\bar{\bm\lambda}_{t}$ and $\bar{\bm\mu}_{t}$ with stochastic gradient ascent iterations as in Figure \ref{fig_training_iteration_shared}. The tweak is that we then keep track of shared multipliers $\bar{\bm\lambda}_{t}$ and $\bar{\bm\mu}_{t}$ and multiplier deviations $\Delta\bm\lambda_{t}(\bfr)$ and $\Delta\bm\mu_{t}(\bfr)$ but not of pointwise multipliers $\bm\lambda_{t} (\bfr)$ and $\bm\mu_{t}(\bfr)$. 

To explain this better refer to the training iteration for hybrid multipliers summarized in Figure \ref{fig_training_iteration_hybrid}. In the input step (I) we evaluate pointwise multipliers as sums of shared multipliers and multiplier deviations [cf. \eqref{eqn_hybrid_multipliers}]. This is done for all multipliers in the primal and dual batches $B_t$ and $C_t$. Having evaluated pointwise multipliers we proceed to the primal step (P) where we update the primal parameter $\bfA_t$ with a stochastic gradient descent iteration on the Lagrangian associated to the problem with pointwise constraints [cf. \eqref{eqn_lagrangian_gradient_descent}]. We then consider two dual updates (D$_{\text{P}}$) and (D$_{\text{S}}$) in which we perform updates of the pointwise [cf. \eqref{eqn_supergradient_ascent_stochastic}] and shared [cf. \eqref{eqn_supergradient_ascent_stochastic_shared}] dual variables, respectively. We close the iteration with the computation of the dual variable deviations [cf. \eqref{eqn_hybrid_multipliers}] in the output step (O).

The hybrid training algorithm is most similar to the pointwise training algorithm. To see this, consider a multiplier that is drawn in batch $C_t$ and again in batch $C_{t+1}$. In this case the computation of shared multipliers is superfluous because the combination of step (O) of iteration $t$ and step (I) of iteration $t+1$ is to recover the pointwise multipliers determined at step (D$_{\text{P}}$) of iteration $t$. The idea of keeping track of shared multipliers is that pointwise multipliers $(\bm\lambda_{t} (\bfr), \bm\mu_{t}(\bfr))$ and $(\bm\lambda_{t} (\bfr'), \bm\mu_{t}(\bfr'))$ associated with different demand vector realizations $\bfr$ and $\bfr'$ are likely related. Thus, we expect the shared update in step (D$_{\text{S}}$) to help with reducing the complexity of finding optimal pointwise multipliers.


\begin{remark} \label{rmk_heterogeneity}\normalfont

The choice to work with pointwise, shared, or hybrid multipliers depends on the heterogeneity of the demand distribution $\bm\rho$. We prefer pointwise multipliers in heterogeneous datasets and shared multipliers in homogeneous datasets. Hybrid multipliers are an intermediate compromise. This is true because optimal Lagrange multipliers $(\bm\lambda^\dagger,\bm\mu^\dagger)$ are subgradients of the constraint perturbation function \cite[Ch. 5]{boyd2004convex}. In particular, $(\bm\lambda^\dagger(\bfr),\bm\mu^\dagger(\bfr))$ defines the sensitivity of the objective function with respect to perturbations of the constraints associated with realization $\bfr$. If realizations $\bfr$ are drawn from a homogeneous distribution we expect similar sensitivity for different realizations of $\bfr$. This translates into similar optimal multipliers $(\bm\lambda^\dagger(\bfr),\bm\mu^\dagger(\bfr))$ for all realizations $\bfr$ which implies that working with shared multipliers carries a small optimality penalty. I.e., the optimal solutions of \eqref{eqn_opf_learning_pointwise} and \eqref{eqn_opf_learning_average_with_penalty} are close. In heterogeneous distributions $\bm\rho$ we expect sensitivity with respect to the perturbation of constraints of different realizations to be very different. This translates to large differences across multipliers $(\bm\lambda^\dagger(\bfr),\bm\mu^\dagger(\bfr))$ associated with different realizations.

\end{remark}


\begin{remark} \normalfont

Solving the pointwise dual problem requires storing significant memory. We need to store $8N+2M$ multipliers for each instance in the dataset. To represent the size and complexity of the entire European high-voltage transmission network, we need approximately $N=13,000$ buses and $M=20,000$ branches~\cite{Josz16-ACPowerFlow} and in our numerical experiments, we found that datasets with $10,000$ instances are practical. In this continental-scale power system, storing all these multipliers requires in the order of 6GB of memory. Although this has been pointed as an unworkable limitation of pointwise constraints in prior contributions~\cite{Park23-Selfsupervised}, we found that these memory requirements are significant but not prohibitive. 

\end{remark}

\section{Experiments}\label{ch4:sec:experiments}

\begin{figure*}

\centering

\small

\renewcommand{\arraystretch}{1.3}

\setlength{\tabcolsep}{3.81mm}

\begin{tabular}
{ 
    l
    l 
    >{\raggedleft\arraybackslash}p{0.8cm}
    >{\raggedleft\arraybackslash}p{0.8cm}
    >{\raggedleft\arraybackslash}p{0.8cm}
    >{\raggedleft\arraybackslash}p{0.8cm}
    >{\raggedleft\arraybackslash}p{0.8cm} 
    >{\raggedleft\arraybackslash}p{1.1cm}
    >{\raggedleft\arraybackslash}p{1.1cm}
    >{\raggedleft\arraybackslash}p{1.1cm}  
}

\midrule

&& \multicolumn{2}{c}{Optimality Gap (\%)} 
&  \multicolumn{2}{c}{Mean Violation (\%)} 
&  \multicolumn{4}{c}{Max Violation  (\%)} \\

\cmidrule(lr){3-4} \cmidrule(lr){5-6} \cmidrule(lr){7-10}
 
&& Mean & Std & Mean & Std & Mean & Std & P95 & Max \\

\midrule

\multirow[c]{5}{*}{IEEE 30} 
 & Dual-P   & 0.18 & 1.41 & 0.49 & 0.62 & 1.83 & 2.15 & 5.65 & 19.94 \\
 & Dual-H   & -0.08 & 0.55 & 1.05 & 0.30 & 5.61 & 0.27 & 6.03 & 6.65 \\
 & Dual-S   & 2.18 & 1.25 & 1.34 & 0.30 & 8.22 & 1.98 & 11.61 & 13.47 \\
 & MSE      & 0.24 & 0.21 & 0.35 & 0.34 & 1.63 & 1.78 & 5.22 & 11.28 \\
 & MSE+Pen. & 0.70 & 0.59 & 0.22 & 0.23 & 1.05 & 0.86 & 3.01 & 7.82 \\
 
\midrule

\multirow[c]{5}{*}{IEEE 57} 
 & Dual-P   & 0.26 & 0.15 & 0.20 & 0.21 & 0.98 & 1.24 & 3.14 & 12.57 \\
 & Dual-H   & 0.33 & 0.17 & 0.71 & 0.53 & 5.20 & 4.29 & 13.27 & 18.51 \\
 & Dual-S   & 0.30 & 0.19 & 0.71 & 0.92 & 3.36 & 4.10 & 11.37 & 23.64 \\
 & MSE      & 0.27 & 0.10 & 0.32 & 0.28 & 1.45 & 1.52 & 3.28 & 22.83 \\
 & MSE+Pen. & 0.19 & 0.18 & 0.22 & 0.35 & 1.09 & 1.69 & 4.09 & 16.71 \\

\midrule

\multirow[c]{5}{*}{IEEE 118} 
 & Dual-P   & -0.18 & 0.03 & 0.29 & 0.14 & 2.27 & 0.43 & 2.86 & 5.44 \\
 & Dual-H   & 0.17 & 0.05 & 0.59 & 0.19 & 4.94 & 0.79 & 6.12 & 7.81 \\
 & Dual-S   & -0.03 & 0.12 & 0.51 & 0.49 & 3.07 & 1.69 & 6.52 & 10.71 \\
 & MSE      & 0.02 & 0.04 & 0.73 & 0.19 & 6.48 & 2.76 & 11.86 & 22.89 \\
 & MSE+Pen. & 0.22 & 0.04 & 0.57 & 0.17 & 3.37 & 1.05 & 4.99 & 9.14 \\

\midrule

\multirow[c]{5}{*}{GOC 179} 
 & Dual-P   & 3.32 & 0.71 & 0.33 & 0.08 & 3.98 & 1.76 & 7.94 & 9.93 \\
 & Dual-H   & 4.80 & 0.93 & 0.36 & 0.09 & 6.37 & 3.30 & 13.05 & 16.07 \\
 & Dual-S   & 4.72 & 0.99 & 0.62 & 0.32 & 5.51 & 3.42 & 11.69 & 13.66 \\
 & MSE      & -0.05 & 0.63 & 2.58 & 1.55 & 9.09 & 6.60 & 19.31 & 145.10 \\
 & MSE+Pen. & 0.25 & 0.64 & 2.06 & 1.32 & 8.68 & 4.54 & 18.32 & 31.46 \\

\midrule

\multirow[c]{5}{*}{IEEE 300} 
 & Dual-P   & 1.80 & 1.76 & 2.15 & 1.39 & 19.34 & 11.98 & 36.93 & 247.38 \\
 & Dual-H   & 1.65 & 1.47 & 1.80 & 0.78 & 31.99 & 24.97 & 85.55 & 169.00 \\
 & Dual-S   & 0.23 & 2.13 & 45.92 & 26.34 & 2,132.19 & 1,779.66 & 5,625.48 & 9,575.20 \\
 & MSE      & 0.60 & 1.22 & 8.00 & 1.58 & 76.19 & 63.57 & 147.20 & 858.42 \\
 & MSE+Pen. & 2.51 & 1.40 & 38.43 & 3.28 & 952.47 & 179.91 & 1251.43 & 1872.07 \\

\midrule

\end{tabular}

\caption{Optimality and Constraint Violation Disaggregated by Network. The optimality gap, mean violation, and maximum violation are computed for each sample in the test sets and then aggregated across all samples. We report several metrics, including the mean, standard deviation, maximum, and 95th percentile. The statistics are computed across all samples grouped by power system. See Section~\ref{sec_evaluation} for details.}

\label{ch4:tab:case_summary}

\end{figure*}


We evaluate the performance of the constrained learning methods of Section \ref{ch4:sec:primal-dual} and two supervised training methods discussed in \ref{ch4:sec:supervised}. In discussions, tables and figures we use the following abbreviations: 

\begin{list}{}{
\setlength{\labelwidth}{18pt}%
\setlength{\labelsep}{2pt}%
\setlength{\leftmargin}{20pt}%
\setlength{\topsep}{5pt}
\setlength{\itemsep}{5pt}}

\item[\textbf{(Dual-P)}] Training with pointwise constraints as summarized in Figure \ref{fig_training_iteration} and discussed in Section \ref{sec_dual_training}.

\item[\textbf{(Dual-H)}] Training with hybrid multipliers as summarized in Figure \ref{fig_training_iteration_hybrid} and discussed in Section \ref{sec_hybrid_multipliers}.

\item[\textbf{(Dual-S)}] Training with shared multipliers as summarized in Figure \ref{fig_training_iteration_shared} and discussed in Section \ref{sec_shared_multipliers}.

\item[\textbf{(MSE)}] Supervised training that imitates the solution of OPF as given by \eqref{eqn_opf_learning_supervised} in \ref{ch4:sec:supervised}.

\item[\textbf{(MSE + Pen.)}] Supervised training with penalties to penalize constraint violations as given by \eqref{eqn_opf_learning_supervised} in \ref{ch4:sec:supervised}.

\end{list}

\noindent We evaluate these five methods in five power systems provided in the Power Grid Library (PGLIB-OPF) \cite[ver. 21.07]{Babaeinejadsarookolaee21-Power}. They are the IEEE 30, IEEE 57, IEEE 118, GOC 179, and IEEE 300 power systems, which, as their names indicate, consist of power networks with  30, 57, 118, 179, and 300 buses, respectively. Each of these networks is provided with a set of reference loads $\bfr$. We generate training and test loads by perturbing reference loads. These loads are chosen uniformly at random with maximum relative perturbations of $\pm 20\%$; see \ref{sec_experimental_details} for details. Perturbation of reference loads is standard practice; see e.g. \cite{Guha09-MachineLearningAC, Zhou23-DeepOPFFTOneDeep, Huang21-DeepOPFNGTFast}.

In all experiments we use graph attention (GAT) networks for the learning parameterization $\Phi(\bfr; \bfA, \bbW )$; see \ref{ch4:sec:gat}. We learn different parameters for each power system but use the same configuration for the GAT network; see \ref{sec_gat_implementation_parameters}. 

Solutions are evaluated in terms of their optimality gap and their constraint violations. Optimality gaps are computed as percentages with respect to the true optimal value. Constraint violations are expressed as percentages of the range of the variable of interest; see \ref{sec_evaluation}. All reported results are for test loads $\bfr$ that are not seeing during training. 

We refer the interested reader to \ref{sec_experimental_details} for further experimental details, including the choice of hyperparameters (\ref{sec_hyperparameters}) and the specific algorithms used to implement the primal, dual, and supervised training loops (\ref{sec_training}).


\begin{figure*}

\centering
        
\small

\renewcommand{\arraystretch}{1.3}

\setlength{\tabcolsep}{3.1mm}

\begin{tabular}
{ 
    l
    l 
    >{\raggedleft\arraybackslash}p{0.8cm}
    >{\raggedleft\arraybackslash}p{0.8cm}
    >{\raggedleft\arraybackslash}p{0.8cm}
    >{\raggedleft\arraybackslash}p{0.8cm}
    >{\raggedleft\arraybackslash}p{0.8cm} 
    >{\raggedleft\arraybackslash}p{0.8cm}
    >{\raggedleft\arraybackslash}p{0.8cm}
    >{\raggedleft\arraybackslash}p{0.8cm}  
    >{\raggedleft\arraybackslash}p{0.8cm}  
    >{\raggedleft\arraybackslash}p{0.8cm}          
}

\midrule

&& \multicolumn{2}{c}{$\Re(\bfs^g)$ (\%)} 
&  \multicolumn{2}{c}{$\Im(\bfs^g)$ (\%)} 
&  \multicolumn{2}{c}{$\lvert V \rvert$ (\%)} 
&  \multicolumn{2}{c}{$\lvert \bff_{f} \rvert$ (\%)} 
&  \multicolumn{2}{c}{$\lvert \bff_{t} \rvert$ (\%)} \\

\cmidrule(lr){3-4} \cmidrule(lr){5-6} \cmidrule(lr){7-8} \cmidrule(lr){9-10} \cmidrule(lr){11-12}
 
&& Mean & Max & Mean & Max & Mean & Max & Mean & Max & Mean & Max \\

\midrule

\multirow[c]{5}{*}{IEEE 30} 
 & Dual-P & 0.20 & 0.25 & 0.69 & 0.70 & 0.74 & 0.85 & 0.68 & 0.68 & 0.15 & 0.15 \\
 & Dual-H & 0.35 & 0.72 & 0.90 & 0.92 & 3.76 & 5.61 & 0.25 & 0.25 & 0.00 & 0.00 \\
 & Dual-S & 0.55 & 0.86 & 0.14 & 0.14 & 6.00 & 8.21 & 0.00 & 0.00 & 0.00 & 0.00 \\
 & MSE & 0.24 & 0.33 & 1.36 & 1.46 & 0.13 & 0.14 & 0.03 & 0.03 & 0.00 & 0.00 \\
 & MSE+Pen. & 0.41 & 0.67 & 0.54 & 0.57 & 0.08 & 0.08 & 0.08 & 0.08 & 0.00 & 0.00 \\

\midrule

\multirow[c]{5}{*}{IEEE 57}
 & Dual-P & 0.13 & 0.15 & 0.52 & 0.61 & 0.36 & 0.45 & 0.00 & 0.00 & 0.00 & 0.00 \\
 & Dual-H & 0.12 & 0.13 & 0.02 & 0.02 & 3.42 & 5.17 & 0.00 & 0.00 & 0.00 & 0.00 \\
 & Dual-S & 0.04 & 0.04 & 2.32 & 2.72 & 1.22 & 1.54 & 0.00 & 0.00 & 0.00 & 0.00 \\
 & MSE & 0.24 & 0.25 & 1.24 & 1.39 & 0.10 & 0.10 & 0.00 & 0.00 & 0.00 & 0.00 \\
 & MSE+Pen. & 0.09 & 0.09 & 0.65 & 0.75 & 0.38 & 0.44 & 0.00 & 0.00 & 0.00 & 0.00 \\

\midrule

\multirow[c]{5}{*}{IEEE 118}
 & Dual-P & 0.33 & 2.20 & 0.52 & 0.64 & 0.59 & 1.12 & 0.00 & 0.00 & 0.00 & 0.00 \\
 & Dual-H & 0.70 & 4.94 & 0.47 & 0.52 & 0.03 & 0.03 & 1.68 & 1.68 & 0.06 & 0.06 \\
 & Dual-S & 0.38 & 2.05 & 0.19 & 0.21 & 0.01 & 0.01 & 0.11 & 0.11 & 1.84 & 2.06 \\
 & MSE & 0.28 & 1.11 & 2.34 & 6.48 & 0.34 & 0.55 & 0.00 & 0.00 & 0.67 & 0.67 \\
 & MSE+Pen. & 0.32 & 1.27 & 1.67 & 3.18 & 0.50 & 0.87 & 0.00 & 0.00 & 0.34 & 0.34 \\

\midrule
\multirow[c]{5}{*}{GOC 179}
 & Dual-P & 0.08 & 0.09 & 0.00 & 0.00 & 1.33 & 3.55 & 0.00 & 0.00 & 0.00 & 0.00 \\
 & Dual-H & 0.05 & 0.06 & 0.00 & 0.00 & 1.75 & 6.37 & 0.00 & 0.00 & 0.00 & 0.00 \\
 & Dual-S & 0.13 & 0.17 & 0.00 & 0.00 & 2.95 & 5.50 & 0.00 & 0.00 & 0.00 & 0.00 \\
 & MSE & 1.33 & 2.12 & 2.16 & 2.52 & 3.44 & 6.79 & 2.95 & 3.92 & 3.02 & 5.35 \\
 & MSE+Pen. & 1.53 & 2.16 & 0.82 & 0.93 & 3.56 & 7.17 & 1.95 & 2.63 & 2.46 & 3.62 \\

\midrule

\multirow[c]{5}{*}{IEEE 300}
 & Dual-P & 0.26 & 0.51 & 4.21 & 9.49 & 5.10 & 16.16 & 0.53 & 0.64 & 0.65 & 0.65 \\
 & Dual-H & 0.23 & 0.46 & 4.18 & 26.09 & 4.15 & 17.01 & 0.32 & 0.34 & 0.14 & 0.15 \\
 & Dual-S & 5.29 & 53.27 & 160.23 & 2132.19 & 18.03 & 55.82 & 22.01 & 43.31 & 24.05 & 48.18 \\
 & MSE & 1.05 & 14.07 & 8.09 & 76.19 & 0.98 & 3.59 & 12.81 & 14.09 & 17.07 & 19.73 \\
 & MSE+Pen. & 0.53 & 1.33 & 75.52 & 943.52 & 114.20 & 611.50 & 1.56 & 2.56 & 0.35 & 0.36 \\

\midrule

\end{tabular}

\caption{Optimality and Constraint Violation Disaggregated by Network and Constraint Type. The table reports the mean of means and maximums of constraint violations for each training methodology and power system. The mean and maximum constraint violations are computed for each sample in the test set and then averaged across all samples. See Section~\ref{sec_evaluation} and Figure~\ref{fig_OPF} for the definition of each variable and constraints. Voltage angle constraints are omitted from the table because we do not observe violations. Equality constraints do not appear in the table since they are substituted out, as explained in Section~\ref{sec_evaluation}}

\label{ch4:tab:constraint_breakdown}

\end{figure*}


\subsection{Feasibility and Optimality Tradeoffs}\label{sec_feasibility_experiments}

Since training with pointwise constraints is introduced as a method to reduce constraint violations and constraint violations are more difficult to eliminate in large scale networks -- which have more constraints -- and corner cases -- by which we mean demand realizations for which constraints are more difficult to satisfy --  we put forth the following hypotheses:

\begin{list}{}{
\setlength{\labelwidth}{18pt}%
\setlength{\labelsep}{2pt}%
\setlength{\leftmargin}{20pt}%
\setlength{\topsep}{5pt}
\setlength{\itemsep}{5pt}}

\item[\textbf{(H1)}] Training with pointwise constraints results in learned parameterizations with smaller constraint violations; possibly at the cost of a reduction in optimality. 

\item[\textbf{(H2)}] Reductions in the violation of constraints are more marked in power systems with larger numbers of buses.

\item[\textbf{(H3)}] Reductions in the violation of constraints are more marked in corner cases. 

\end{list}

\begin{figure*}
    \centering
    \includegraphics[width=\linewidth, height=0.8\textheight, keepaspectratio]{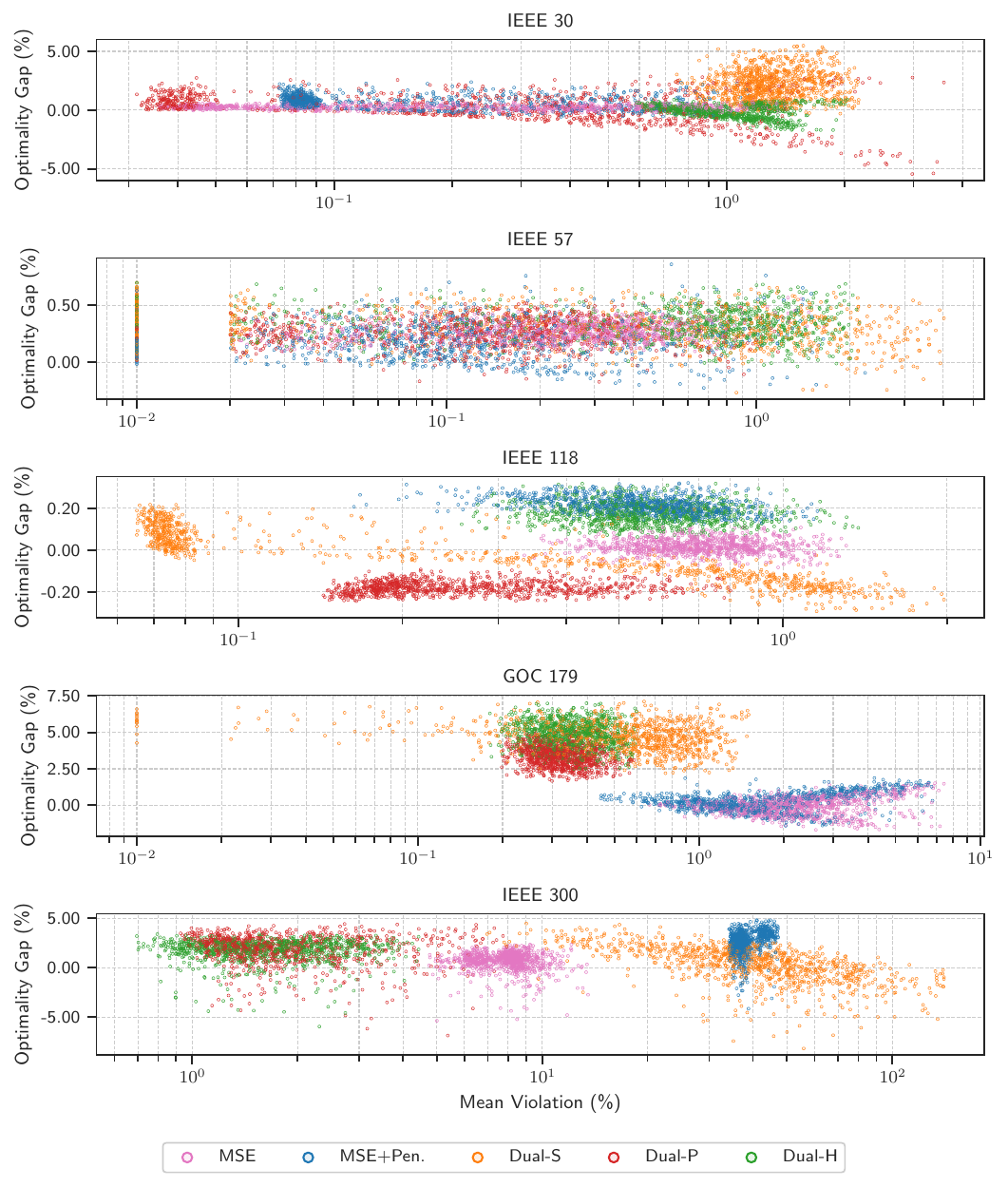}
    \caption{
        The relationship between optimality gap and mean constraint violation within each test sample is grouped by power system and training method.
        Each point represents a test sample with different hues corresponding to different training methods.
        Note the different x-axis and y-axis scales for the different power systems.}
    \label{ch4:fig:tradeoff_mean}
\end{figure*}

\begin{figure*}
    \centering
    \includegraphics[width=\linewidth, height=0.8\textheight, keepaspectratio]{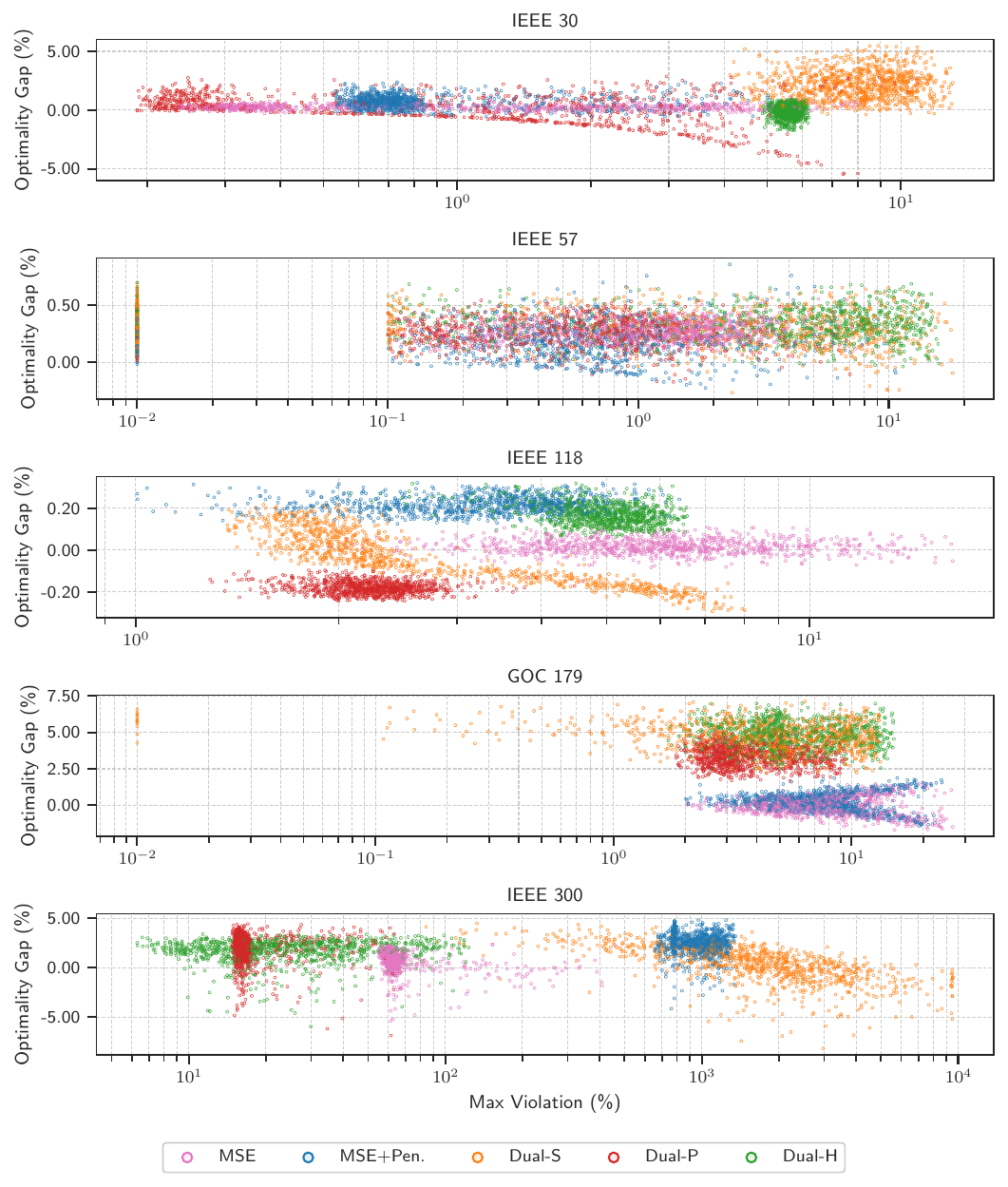}
    \caption{
        Shows the relationship between optimality gap and maximum constraint violation within each test sample, grouped by power system and training method.
        Each point represents a test sample with different hues corresponding to different training methods.
        Note the different x-axis and y-axis scales for the different power systems.}
    \label{ch4:fig:tradeoff_max}
\end{figure*}

\noindent Hypotheses (H1)-(H3) are largely validated by numerical experiments. The tables in Figures \ref{ch4:tab:model_summary}, \ref{ch4:tab:case_summary} and \ref{ch4:tab:constraint_breakdown} as well as Figures \ref{ch4:fig:tradeoff_mean} and \ref{ch4:fig:tradeoff_max} show that constraint violations are smaller for (Dual-P) than they are for (Dual-S), (MSE) and (MSE + Pen.). Often but not always, smaller constraint violations come at the cost of a larger optimality gap. These two statements validate Hypothesis (H1). 

The tables in Figures \ref{ch4:tab:case_summary} and \ref{ch4:tab:constraint_breakdown} as well as Figures \ref{ch4:fig:tradeoff_mean} and \ref{ch4:fig:tradeoff_max} show that improvements in constraint violations are more marked in larger networks. In fact, we see little advantage to the use of pointwise constraints in the small IEEE 30 and IEEE 57 power systems. We see that (Dual-P) outperforms (Dual-S), (MSE) and (MSE + Pen.) in the larger IEEE 118, GOC 179 and IEEE 300 systems. Improvements are particularly marked in the IEEE 300 system. These statements validate Hypothesis (H2).

Figures \ref{ch4:fig:tradeoff_mean} and \ref{ch4:fig:tradeoff_max} are scatter plots of optimality and constraint violation statistics for \emph{individual} realizations of the power demand vector $\bfr$. We see that the cloud of (Dual-P) realizations for the IEEE 118, GOC 179 and IEEE 300 power systems are not only shifted towards smaller constraints but are also more concentrated. An extreme example of this behavior is the IEEE 118 network in which (Dual-S) actually works better on average but has a long tail of realizations with large constraint violations. These statements validate Hypothesis (H3).

Further recall that training with hybrid multipliers maintains the use of individual multipliers for each realization of the power demand vector $\bfr$. We therefore expect that training with hybrid multipliers is similar to training with pointwise constraints:

\begin{list}{}{
\setlength{\labelwidth}{18pt}%
\setlength{\labelsep}{2pt}%
\setlength{\leftmargin}{20pt}%
\setlength{\topsep}{5pt}
\setlength{\itemsep}{5pt}}

\item[\textbf{(H4)}] Training with hybrid multipliers yields similar outcomes to training with hybrid multipliers. 

\end{list}

\noindent Hypothesis (H4) is disproved by experiments. We did not succeed in training (Dual-H) to replicate the performance of (Dual-P).


\subsection{Detailed Discussion of Experimental Outcomes}

Table~\ref{ch4:tab:model_summary} summarizes the aggregate performance of each training method across all power systems.
In aggregate, the table shows that the supervised methods perform better in optimality gap, while the primal-dual methods perform better in terms of mean and maximum violations.
In terms of the optimality gap, there is a clear distinction between the supervised and primal-dual methods.
The MSE and MSE+Penalty methods have an average optimality gap of $0.22\pm0.66\%$ and $0.77\pm1.16\%$, respectively.
Meanwhile, the Dual-S, Dual-P, and Dual-H methods methods have an optimality gap of $1.48\pm2.16\%$, $1.95\pm3.17\%$, and $1.38\pm1.99\%$, respectively.
In terms of the mean violations, the methods achieve an average relative violation of $2.40\pm3.09\%$ and $8.30\pm15.17\%$ for the supervised methods and $9.82\pm21.56\%$, $0.87\pm1.01\%$, and $0.90\pm0.68\%$ for the primal-dual methods.
The differences in terms of the maximum violations are even more pronounced.

Notably, the MSE+Penalty and Dual-S methods have a significantly larger constraint violation in Table~\ref{ch4:tab:model_summary}, which is particularly unexpected for the MSE+Penalty method.
We would expect that adding a penalty term to the loss function would reduce the constraint violations, not exacerbate them.
To understand this phenomenon, consider Table~\ref{ch4:tab:case_summary}, which provides a detailed breakdown of the constraint violations, grouped by power system.
Across almost all power systems, incorporating constraint penalties reduces the frequency and severity of constraint violations over just using the MSE loss.
This is expected and comes with a trade-off in optimality gap.
Exceptionally, for the IEEE 300 power system, the MSE+Penalty method produces highly infeasible solutions.
Similarly, Table~\ref{ch4:tab:case_summary} shows that the Dual-S method has large violations for the IEEE 300 power system.

Table~\ref{ch4:tab:constraint_breakdown} further breaks down constraint violations by variable type.
The table highlights one side-effect of substituting the power flow equality constraints in \eqref{eqn_branch_from_elementwise} and \eqref{eqn_branch_to_elementwise}. When equality constraints are eliminated, errors are introduced in some of the other inequality constraints. Since we solve for the voltages using the power injections from the generator buses during the substitution process, the inequality constraints on power generation are unaffected. This is well illustrated by the violations of MSE+Penalty and Dual-S methods on the IEEE 300 bus system in Table~\ref{ch4:tab:constraint_breakdown}. The violations of real generation limits are low, and we see large violations of the reactive generation limits and the voltage magnitude limits instead.

Figures~\ref{ch4:fig:tradeoff_mean} and~\ref{ch4:fig:tradeoff_max} visualize the relationship between optimality and feasibility.
Figure~\ref{ch4:fig:tradeoff_mean} plots the relationship between optimality gap and the mean constraint violation across all samples in the test set.
Similarly, Figure~\ref{ch4:fig:tradeoff_max} relates optimality gap and the maximum constraint violation.

In Figures~\ref{ch4:fig:tradeoff_mean} and~\ref{ch4:fig:tradeoff_max}, the Dual-S method stands out as having the widest distribution.
In Dual-S, the Lagrange multipliers are shared between samples in the dataset.
Therefore, outside of the augmented Lagrangian term, there is no mechanism to learn constraints on individual samples.
With shared multipliers, we can only learn to satisfy constraints in expectation.
This results in Dual-S having the most variability between samples in the test set and is most clearly visible in the IEEE 118 test case in Figure~\ref{ch4:fig:tradeoff_mean}, where the Dual-S method has some samples with the lowest violations of all methods, but also has some samples with the highest violations.

For the IEEE 30 and 57 bus systems, Figures~\ref{ch4:fig:tradeoff_mean} and~\ref{ch4:fig:tradeoff_max} clearly show that the Dual-S and Dual-H have the highest violations, without a substantial difference in optimality gap.
For these power systems, Table~\ref{ch4:tab:case_summary} offers a better medium of comparison between the MSE, MSE+Penalty, Dual-P methods.
On the IEEE 30 bus system, the Dual-P method has the lowest cost of all methods but also the highest violations. Moreover, the Dual-P method has the highest standard deviation of its metrics.
On the IEEE 57 bus system, the three methods are very similar across all metrics.
Overall, for the IEEE 30 and 57 bus systems, the differences between the MSE, MSE+Penalty, and Dual-P methods are minor.

On the IEEE 118 bus system, the Dual-P method performs best.
Figure~\ref{ch4:fig:tradeoff_max} shows that the Dual-P method offers the lowest maximum violations at the lowest cost of all methods.
Comparison using the trade-off between optimality gap and mean violations is more nuanced.
In Figure~\ref{ch4:fig:tradeoff_mean}, the Dual-S method has a cluster of samples with sub 0.1\% average violations for the 118 bus system. However, as discussed above, Dual-S has a large spread in violations.

The dual methods all perform particularly poorly on the GOC 179 bus system.
Figure~\ref{ch4:fig:tradeoff_max} shows that, in terms of maximum violations, the dual methods perform similarly to the supervised methods but have a significantly higher optimality gap.
In terms of mean violations, as shown in Figure~\ref{ch4:fig:tradeoff_mean}, the dual methods have some advantages over the supervised methods.
The large difference between the mean and maximum violations is accounted for by Table~\ref{ch4:tab:constraint_breakdown}, which shows the dual methods satisfy all the constraints except for the voltage magnitude constraints.
We qualitatively observed that for the GOC 179 bus system, the gradient ascent descent method did not converge during training, which is why the performance of the dual methods is so much worse.
We hypothesize that the choice of dual step size was too high. A small dual step size is required for the theoretical convergence of the gradient ascent descent method on nonconvex-concave min-max problems \cite{Lin20-GradientDescentAscent}.

On the IEEE 300 bus system, the dual methods with pointwise multipliers appear to have the best performance.
As discussed above, the MSE+Penalty and the Dual-S method fail on this system.
The MSE method has the lowest average optimality gap at $0.60\pm1.22$, while the Dual-P and Dual-H methods have optimality gaps of $1.80\pm1.76\%$ and $1.65\pm1.47\%$, respectively.
However, the model trained using the MSE method performs much worse in terms of feasibility -- an average mean violation of $8.00\pm1.58\%$ as opposed to $2.15\pm1.40\%$ and $1.80\pm0.78\%$.
The differences in performance are also clearly visible in Figures~\ref{ch4:fig:tradeoff_mean} and~\ref{ch4:fig:tradeoff_max}.
The Dual-H method has some samples with lower maximum violations, but the mean and the spread of the maximum violations are higher.

Overall, the Dual-P method appears to be the best-performing method regarding the trade-off between optimality and feasibility.
Even though the hyperparameter search was explicitly performed to favor the supervised methods on the IEEE 118 bus system, the Dual-P method performs best.
On smaller systems, the Dual-P performs similarly to the supervised methods with minor differences, while the Dual-H and Dual-S methods tend to perform worse.
The Dual-P method performs poorly on the GOC 179 system, but with Dual-H, it stands out with the best performance on the IEEE 300 bus system.

\section{Conclusions}\label{ch4:sec:conclusions}
We have argued that the Lagrangian dual formulation used for constrained learning on optimal power flow is flawed.
Instead, we propose a pointwise formulation and hypothesize that it would produce more feasible solutions.
Additionally, we introduce a graph attention network that fully represents all the variables in the OPF problem.
Using this proposed architecture we compare several supervised and primal-dual methods on a variety of power systems.

The experimental results demonstrated that while supervised methods generally achieve better optimality gaps, primal-dual methods with pointwise Lagrange multipliers consistently produce more feasible solutions with significantly lower constraint violations. The hybrid approach of combining shared and pointwise multipliers also showed promising results, particularly on larger power systems.

The performance of different methods varied significantly across power system sizes. On smaller systems, the differences between methods were minor, with all approaches achieving comparable optimality and feasibility. However, on larger systems, particularly the IEEE 300 bus system, the advantages of the pointwise primal-dual approach became more apparent. The pointwise method successfully balanced optimality and feasibility on the largest test case. In contrast, both the supervised penalty and the shared multiplier approach failed to produce feasible solutions. These results suggest that the pointwise primal-dual method is better suited for large-scale power systems.

However, none of the tested methods reliably produced feasible solutions on the larger datasets. This could be a limitation of the GNN architecture rather than an issue with the training methods.
Given the overall better performance of the pointwise dual method, future work could compare the impact of the choice of architecture.
Improving the convergence rate of the dual methods could also enhance their performance. 
Lowering the step size could alleviate convergence issues, but this could also lead to longer training times.
Further work is also needed to study more complex and general problem formulations. This includes using realistic demand vector distributions and removing the assumption that cost functions are constant. 

\bibliographystyle{IEEEtran}
\bibliography{IEEEabrv,settings,opf}

\setcounter{section}{0}
\renewcommand{\thesection}{Appendix \Roman{section}}

\section {Supervised Learning} \label{ch4:sec:supervised}

.


In \eqref{eqn_opf_learning_average} and \eqref{eqn_opf_learning_pointwise} the learning parameterization $\Phi(\bfr; \bfA, \bbW )$ is substituted directly into the OPF problem \eqref{eqn_opf_problem}. Alternatively, we can consider solutions of \eqref{eqn_opf_problem} and train parameters that imitate these optimal solutions,
\begin{alignat}{3}\label{eqn_opf_learning_supervised}
    \bfB^{\dagger} =~ 
    &  \underset{\bfA}{\text{argmin}} ~
    && \bbE_{\bm\rho} \Big[\,
            \big \|\, \Phi(\bfr; \bfA, \bbW )  
                 - [\bfs^\dagger (\bfr), \bfv^\dagger (\bfr)]  \, \big\| \, \Big] .
\end{alignat}
This is a supervised learning approach to estimating solutions of OPF \cite{Guha09-MachineLearningAC,Owerko20-OptimalPowerFlow}. Since it utilizes a mean squared error (MSE) loss, we call 
\eqref{eqn_opf_learning_supervised} the MSE OPF problem.

Asides from incurring the computational cost of solving \eqref{eqn_opf_problem}, supervised learning approaches tend to yield power and voltage estimates $[\bfs,\bfv] = \Phi(\bfr; \bfA^{\dagger}, \bbW )$ with significant constraint violations. Constraint violation in supervised learning can be mitigated by adding constraint satisfaction penalties \cite{Huang22-DeepOPFVSolvingACOPF, Pan21-DeepOPFDeepNeural, Pan23-DeepOPF, Zhou23-DeepOPFFTOneDeep}. . This can be done with the same penalty $P (\Phi(\bfr; \bfA, \bbW ))$ that we define \eqref{eqn_opf_augmentation_penalty} to use in \eqref{eqn_opf_learning_pointwise},
\begin{alignat}{3}\label{eqn_opf_learning_supervised_penalties}
    \bfB^{\ddagger} =~ 
    &  \underset{\bfA}{\text{argmin}} ~
    && \bbE_{\bm\rho} \Big[\,
            \big \|\, \Phi(\bfr; \bfA, \bbW )  
                 - [\bfs^\dagger (\bfr), \bfv^\dagger (\bfr)]  \, \big\| \, \Big] 
       \nonumber \\ 
    &
    &&   \qquad\qquad +~ P \Big(\Phi(\bfr; \bfA, \bbW )\Big)
\end{alignat}
We refer to \eqref{eqn_opf_learning_supervised_penalties} as the penalized MSE OPF problem. Numerical experiments indicate that \eqref{eqn_opf_learning_supervised_penalties} under-performs constrained learning in terms of constraint satisfaction (Section \ref{ch4:sec:experiments}).

\section{Graph Attention Networks}\label{ch4:sec:gat}

A graph attention (GAT) network receives as inputs a set of node features $\bfx_{i} = \bfx_{i}^0 \in \bbR^{d_n}$ associated with nodes $i\in[1,N]$ and a set of edge features $\bfe_{ij}\bbR^{d_e}$ associated with directed edges $(i,j) \in E$. Node features are processed sequentially by a stack of $L$ layers so that output features $\bfx_{i}^{l-1} \in \bbR^{d_n}$ of layer $l-1$ become input features to layer $l$. The update of \emph{node features} at layer $l$ consists of a linear combination of the node features of neighboring nodes $j$ at the output of the previous layer,
\begin{align}\label{eqn_value_multiplication}
    \bfz_{i}^l 
        ~=~ \bfA_{\text{NF}}^l \times
                \sum_{ j : (i,j) \in E} ~
                    \frac{ \exp (A_{ij}^l ) }
                             { ~\sum_{ k : (i,k) \in E} \exp( A_{ik}^l )~ } ~ 
                                 \bfx_j^{l-1}.
\end{align}
In this expression the node feature update matrix $\bfA_{\text{NF}}^l \in \bbR^{d_n \times d_n}$ is a trainable parameter of layer $l$ and the scalars $A_{ij}^{l}$ are attention coefficients that we will compute to determine how much the neighboring signal $\bfx_j^{l-1}$ influences the update of the feature $\bfx_i^{l-1}$ of node $i$. We convene that $(i,i) \in E$ and emphaisze that the matrix $\bfA_{\text{NF}}^l$ is the same for all nodes $i$.

To compute the attention coefficients $A_{ij}^{l}$, concatenate edge features $\bfx_{ij}$ with the features $ \bfx^{l-1}_i$ and $\bfx^{l-1}_j$ of the corresponding vertices to produce the modified edge feature $\bfA_{\text{EF}}^l [\bfx_{ij}, \bfx^{l-1}_i, \bfx^{l-1}_j] \in \bbR^{d_a}$ as determined by the trainable \emph{edge feature} update matrix $\bfA_{\text{EF}} \in \bbR^{d_a \times (2d_n + d_e)}$. Further introduce a trainable \emph{attention} vector $\bfa_{\text{ATT}}^l \in \bbR^{d_a}$ and let the attention coefficient $A_{ij}^l$ be the inner product
\begin{align}\label{eqn_attention}
    A_{ij}^l 
        ~=~ \Big\langle \,
                \bfa_{\text{ATT}}^l , \,
                    \sigma \Big(\, \bfA_{\text{EF}}^l \, 
                        \big[ \, \bfx_{ij}, \bfx^{l-1}_i, \bfx^{l-1}_j \, \big]  
                            \,\Big) \,\Big\rangle ~.
\end{align}
We emphasize that in the computation of attention coefficients the edge feature matrix  $\bfA_{\text{EF}}^l$ and the attention coefficient $\bfa_{\text{ATT}}^l$ are the same for all edges.

To complete layer $l$, we perform a relative update of the features $\bfx_{i}^{l-1}$

To complete layer $l$ we perform a relative update of the features $\bfx_{i}^{l-1}$. We do that by adding to $\bfx_{i}^{l-1}$ the output of multilayer perceptron (MLP) with input $\bfx_{i}^{l-1}  + \bfz_{i}^l$ and trainable parameters $\bfA^l_{\text{MLP}}$,
\begin{align}\label{eqn_layer_MLP}
    \bfx_{i}^{l} 
        = \bfx_{i}^{l-1} + \Psi_l \Big( \, \bfx_{i}^{l-1}  + \bfz_{i}^l;  \bfA^l_{\text{MLP}} \, \Big). 
\end{align}
As per \eqref{eqn_value_multiplication}-\eqref{eqn_layer_MLP}, layer $l$ of a GAT network takes as input the oputput node features $\bfx^{l-1}_i$ of the previous layer. These outputs are used first to compute the attention coefficients $A_{ij}^l$ defined in \eqref{eqn_attention}. The attention coefficients $A_{ij}^l$ of node $i$ along with node features $\bfx_{i}^{l-1}$ of neighboring nodes $j$ determine the updated feature $\bfz_{i}^l$ as the linear combination in \eqref{eqn_value_multiplication}. The updated node feature $\bfz_{i}^l$ is then passed through the MLP in \eqref{eqn_layer_MLP} to produce $\bfx_{i}^{l}$. This is the output feature of Layer $l$ at node $i$. It is important that the trainable parameters $\bfA^l_{\text{EF}}$ and $\bfa^l_{\text{ATT}}$ are the same for all edges and that the trainable parameters $\bfA^l_{\text{NF}}$ are the same for all nodes. These choices make the attention and linear combination operations in \eqref{eqn_value_multiplication} and \eqref{eqn_attention} equivariant to permutations of the graph \cite{2021_Luana_Ruiz_2008.01767}. Further note that the MLP in \eqref{eqn_layer_MLP} is local to node $i$ and that the the same parameter $\bfA^l_{\text{MLP}}$ is used at all nodes $i$. This operation is therefore independent of the structure of the graph. It follows that layers of GAT networks -- thus, GAT networks themselves -- are equivariant to permutations. This is a fundamental general property of graph neural networks in general and GATs in partcular that allows exploitation of graph symmetries \cite{2021_Luana_Ruiz_2008.01767}. 


\subsection{Graph Attention Networks for Optimal Power Flow}

The set of inputs and outputs of a GAT network matches the structure of an OPF problem if we make the following identifications: (i) Input node features $\bfx_{i}=\bfx_{i}^0$ are power demands $r_i$ and bus parameters $\bfw_i$. (ii) Edge features $\bfx_{ij}$ are branch parameters $\bfw_{ij}$. (iii) Output features $\bfx_i^L$ are power generations $s_i$ and voltages $v_i$ -- see Figures \ref{fig_variables} and \ref{fig_parameters} and Equations \eqref{eqn_bus_features} and \eqref{eqn_branch_features} for the definition of OPF variables and parameters.

Instead of using direct associations of OPF variables and parameters to GAT features, we mediate the associations with MLPs. Start then by processing power demands $r_i$ and bus parameters $\bfw_i$ at node $i$ with an MLP. We call this MLP the \emph{node input} MLP and let $\bfA_\text{NI}$ denote its parameters,
\begin{alignat}{3}\label{eqn_input_MLP}
    &\bfx_{i}  = \bfx_{i}^0 
        && = \Psi_\text{NI} \big(\,[r_i, \bfw_i]; \bfA_\text{NI} \,\big) .
\end{alignat}
The outputs $\bfx_{i} = \bfx_{i}^0$ of these MLPs map to the input node features of the GAT network. I.e. they are used in \eqref{eqn_value_multiplication}-\eqref{eqn_layer_MLP} at the input layer $l=1$ for which $l-1=0$. Observe that, as is the case of \eqref{eqn_layer_MLP}, this operation is local to node $i$ and, therefore, independent of the structure of the graph. 

We continue with the branch features $\bfw_i$ that we map to edge features of the GAT network using the \emph{edge input} MLP. Denoting the parameters of the edge input MLP as $\bfA_\text{EI}$ we write this map as,
\begin{alignat}{3}\label{eqn_edge_input_MLP}
    &\bfx_{ij} 
        && = \Psi_\text{EI}(\bfw_{ij}; \bfA_\text{EI})  .
\end{alignat}
The outputs $\bfx_{ij}$ of these MLPs at different edges are the edge features of the GAT network that we use to compute attention coefficients in \eqref{eqn_attention}. Analogous to \eqref{eqn_layer_MLP} and \eqref{eqn_input_MLP}, this operation is local to \emph{edge} $(i,j)$ and, consequently, independent of the structure of the graph. 

After $L$ GAT layers the oputput $\bfx^L$ is mapped to powers and voltages with the \emph{output} MLP. Using $\bfA_\text{OUT}$ to denote parameters of this MLP, we write
\begin{alignat}{3}\label{eqn_output_MLP}
    &[s_{i}, v_{i} ]  
        && = \Psi_\text{OUT}(\bfx_i^L; \bfA_\text{OUT}) .
\end{alignat}
This MLP is also local to node $i$, thus, independent of the structure of the graph. 

The powers and voltages $[s_{i}, v_{i}]$ are the outputs we use for training in Figures \ref{fig_training_iteration}, \ref{fig_training_iteration_shared} and \ref{fig_training_iteration_hybrid}. I.e., the generic learning parameterization in \eqref{eqn_learning_parameterization} is instantiated to 
\begin{align}\label{eqn_learning_parameterization_with_GAT}
    [\Phi(\bfr; \bfA, \bbW )]_i
        ~=~ [s_{i}, v_{i} ] 
        ~=~ \Psi_\text{OUT}(\bfx_i^L; \bfA_\text{OUT}).
\end{align}
The trainable parameter $\bfA$ groups the parameters $\bfA_{\text{NF}}^l$, $\bfA_{\text{EF}}^l$, $\bfa_{\text{ATT}}^l$ and $\bfA_{\text{MLP}}^l$ of all layers of the GAT network as well as the parameters $\bfA_{\text{NI}}^l$, $\bfA_{\text{EI}}^l$ and $\bfA_{\text{OUT}}^l$ of the node input, edge input, and output MLPs in \eqref{eqn_edge_input_MLP}, \eqref{eqn_output_MLP} and \eqref{eqn_learning_parameterization_with_GAT}.


\subsection{Multilayer Perceptrons}

The OPF GAT as described by \eqref{eqn_value_multiplication}-\eqref{eqn_learning_parameterization_with_GAT} makes use of several multilayer perceptrons MLPs. We represent MLPs with the generic notation $\Psi(\bfy, \bfB)$ where $\bfy$ denotes an input and $\bfB$ a trainable parameter. An MLP is a composition of layers, each of which is itself a composition of a linear map with a pointwise nonlinearity,
\begin{align}\label{eqn_MLP_def}
    \bfy = \sigma\big(\, \bfB_l \bfy^{l-1} \, \big) .
\end{align}
The trainable parameter $\bbB$ groups all of the matrices $\bfB_l$ used at individual layers. We complete the definition of an MLP with the specification of the input and output
\begin{align}\label{eqn_MLP_def_boundaries}
    \bfy^0 = \bfy, \qquad
    \Psi(\bfy, \bfB) = \bfy^L .
\end{align}
The specification of an MLP is given by the number of layers $L$ and the dimensions of all the intermediate outputs $\bfy^l$. The intermediate outputs are called hidden units.

\subsection{Implementation Parameters}\label{sec_gat_implementation_parameters}

In all experiments, we use the same model architecture. The GAT network contains $L = 20$ layers. The dimensions of the input and output node features as well as the dimensions of the edge features are $d_n = d_e = 64$ in all layers. The attention dimension is $d_a=128$ in all layers. The nonlinearity function in the computation of attention coefficients in \eqref{eqn_attention} is a rectified linear unit (ReLU).

All of the MLPs, i.e., the GAT MLP in \eqref{eqn_layer_MLP}, the node input MLP in 
\eqref{eqn_input_MLP}, the edge input MLP in \eqref{eqn_edge_input_MLP} and the output MLP in \eqref{eqn_output_MLP}, have two layers and the dimension of the single hidden unit is $64$. All of these MLPs have different input and output dimensions to match the dimensions of their corresponding inputs and outputs. E.g., the GAT MLP in \eqref{eqn_layer_MLP} has input and output dimensions equal to $d_n = 64$ and the output MLP in \eqref{eqn_output_MLP} has input dimension $d=64$ and output dimension 4 corresponding to the real and imaginary parts of $s_i$ and $v_i$. We found that it improved performance to have output MLPs with different parameters at buses that have a generator and buses that do not. This is a reasonable modification as the generated power must be $s_i=0$ at nodes that do not have a generator attached. All of these MLPs use ReLU nonlinearities in their two layers

\section{Experimental Details}\label{sec_experimental_details}

\subsection{Dataset Generation} \label{sec_dataset}

We generate datasets for each power system by perturbing the reference load.
Each power system provides reference values for the complex power demand at each node, which we denote as $\bfs^d_{\text{ref}} \in \mathbb{R}^{N \times 2}$.
We obtain a dataset by sampling element-wise from a uniform distribution around the reference loads, which is a common practice in literature \cite{Guha09-MachineLearningAC, Zhou23-DeepOPFFTOneDeep, Huang21-DeepOPFNGTFast}.
Load samples are given by:
\begin{equation} \\
    \bfs^d_i \sim \text{Uniform}\left(0.8 \bfs^d_{\text{ref}}, 1.2 \bfs^d_{\text{ref}}\right) \label{ch4:eq:load_distribution}
\end{equation}
for all $i \in \{1,\dots,|\calD|\}$.
The reference load is perturbed by up to 20\% in either direction.
We generate $10,000$ feasible load samples with an 8:1:1 split for training, validation, and testing.
For each load sample, $\bfs^d_i$, we generate ground truth labels $\bfs^g_i$ and $\bfv_i$ using Powermodels.JL \cite{Coffrin18-PowerModelsjl} and the HSL \cite{Rees-HSLCollectionFortran} ma57 solver in IPOPT \cite{Wachter06-Implementation}.
These ground truth labels are necessary for supervised training but not primal-dual methods.

\subsection{Hyperparameter Search}\label{sec_hyperparameters}

We design our experimental methodology to limit the impact of the choice of hyperparameters on our conclusions.
Hyperparameter tuning generally has a sizable impact on deep learning performance and can be a source of publication bias.
In this work, we hypothesized that training with primal-dual methods from Section~\ref{ch4:sec:primal-dual}, especially the pointwise method, is advantageous over the supervised methods from Section~\ref{ch4:sec:supervised}.
Therefore, we specifically tune hyperparameters to maximize the performance of the supervised methods and then use the same hyperparameters for the primal-dual methods.
All the hyperparameter searches were performed on the IEEE 118 dataset.

We use hyperband search \cite{Li17-HyperbandNovel} to tune the hyperparameters for the supervised penalty method (MSE+Penalty).
During the search we vary the latent dimension $d$, number of heads $H$, primal learning rate $\eta_\mathrm{primal}$, primal weight decay $\omega_\mathrm{primal}$, the dropout rate $\alpha$, and the penalty weights $w_h$ and $w_g$, which we restrict to $w_h = w_g$.
The parameters are sampled from the sets indicated in Table~\ref{ch4:tab:hyperparameters_supervised} using a tree-structured Parzen estimator (TPE)~\cite{Bergstra11-Algorithms}.
We use the Optuna~\cite{Akiba19-OptunaNextgeneration} implementation of Hyperband and TPE.
We run 400 trials using hyperband search with a reduction factor of 3, a minimum of 20 epochs per trial, and a maximum of 200 epochs per trial.

We minimize a metric that balances optimality and feasibility throughout the hyperparameter search.
This objective is distinct from the training loss, which cannot be used since it depends on the choice of $w_h$ and $w_g$, which are being searched over.
Instead, we use an \emph{invariant} metric that does not depend on $w_h$ and $w_g$:
\begin{align}
    \calI(\bfs^g_i, \hat{\bfs}^g_i, \hat{\bfv}_i)
     & = \frac{C(\hat{\bfs}^g_i)}{\bar{C}} + 10^3 \frac{\lVert \bm\varepsilon_i^h \rVert_1}{N_h} + 10^3 \frac{\lVert \bm\varepsilon_i^g \rVert_1}{N_g} \label{ch4:eq:invariant_metric} \\
    \bm\varepsilon_i^h
     & = |\bfh(\bfs^d_i, \hat{\bfs}^d_i, \hat{\bfv}_i)| \label{ch4:eq:equality_violation}                                                                                              \\
    \bm\varepsilon_i^g
     & = \left[\bfg(\bfs^g_i, \hat{\bfs}^g_i, \hat{\bfv}_i)\right]_+ \label{ch4:eq:inequality_violation}
\end{align}
where $\bar{C}$ is the average cost of the IPOPT solution on the training set, $\bm\varepsilon_i^h \in \reals^{N_h}$ are equality constraint violations, $\bm\varepsilon_i^g \in \reals^{N_g}$ are inequality constraint violations, $N_h = 2N$ is the number of equality constraints, and $N_g = 6N+2M$ is the number of inequality constraints.
The constants of $10^3$ are chosen to scale the constraint penalty term in the same order as the objective when the average violation error is $10^{-3}$.

\begin{table}
    \centering
    \caption{Hyperparameter search for supervised methods.}
    \label{ch4:tab:hyperparameters_supervised}
    \begin{tabular}{cccc}
        \toprule
        Hyperparameter           & Search Set                & {Best}        & {Importance} \\
        \midrule
        $d$                      & $\{32, 64, 128, 256\}$    & 64            & 0.01         \\
        $H$                      & $\{1, 2, 4, 8\}$          & 2             & 0.03         \\
        $\eta_\mathrm{primal}$   & $[\num{e-5}, \num{e-2}]$  & \num{3.0d-4}  & 0.41         \\
        $\omega_\mathrm{primal}$ & $[\num{e-16}, \num{e-2}]$ & \num{6.9d-15} & 0.22         \\
        $w_h, w_g$               & $[\num{e-2}, \num{e2}]$   & 72            & 0.13         \\
        $\alpha$                 & $[0, 1]$                  & 0.0           & 0.20         \\
        \bottomrule
    \end{tabular}
\end{table}

The parameters of the best trial, in terms of the trade-off between optimality and feasibility, are reported in Table~\ref{ch4:tab:hyperparameters_supervised}.
Additionally, we report the importance of each hyperparameter on a unit scale using fANOVA \cite{Hutter14-EfficientApproach}.
We use the best hyperparameters from Table~\ref{ch4:tab:hyperparameters_supervised} for all training methods where applicable and unless otherwise specified.
However, due to the low importance of $d, H$ on performance, we used $d=128$ and $H=4$ instead of the values from Table~\ref{ch4:tab:hyperparameters_supervised}.
Since the trials in the hyperparameter search lasted 200 epochs, this was a subjective decision to increase capacity, which may benefit performance when training for more epochs.
These parameters are used for training both the supervised (MSE) and the supervised penalty method (MSE+Penalty).

We primarily use the same hyperparameters as above for the primal-dual models, with a minor difference.
Specifically, we use an order of magnitude smaller penalty weight of $w_h = w_g = 5$.
Additionally, we tune the learning rates for the shared and pointwise Lagrange multipliers.
We observed the best performance on the IEEE 118 system with $\eta_\mathrm{sh} = \num{e-2}$ and $\eta_\mathrm{pw} = \num{5e3}$.
To arrive at these values, we trained Dual-S and Dual-P models for 1000 epochs each on the IEEE 118 system, with a variety of learning rates from $\eta_\mathrm{sh} \in [\num{e-4}, \num{e-4}]$ and $\eta_\mathrm{pw} \in [\num{e0}, \num{e6}]$.
We chose learning rates for each model that minimized the objective $\calI$ from Equation~\eqref{ch4:eq:invariant_metric} over the validation dataset.
The Dual-H method uses both learning rates, $\eta_\mathrm{sh}$ and $\eta_\mathrm{pw}$ since the shared and pointwise Lagrange multipliers are used.

\subsection{Training}\label{sec_training}

The supervised models are trained using stochastic gradient descent with hyperparameters from Table~\ref{ch4:tab:hyperparameters_supervised}. The unsupervised models are trained using the algorithms in Figures \ref{fig_training_iteration}, \ref{fig_training_iteration_shared}, and \ref{fig_training_iteration_hybrid}. The generation cost is normalized by the cost weight $w_c = 0.1 \bar{C}^{-1}$, which we set to one-tenth of the inverse of the average cost of the IPOPT solution on the training set. The augmented inequality and equality weights are set to $w_h = w_g = 5$.
For those weights, we chose values lower than those used during supervised training to avoid gradient pathologies found in~\cite{Wang21-Understanding}. The unsupervised models are trained with a supervised aid for the first 500 epochs. During this period, we add the MSE loss in the objective of \eqref{eqn_opf_learning_supervised} to the Lagrangian with an initial weight of 10 that linearly decays to 0 over the aid period. No dual updates are performed for the first 250 epochs.

We use different gradient descent algorithms for GAT model parameters and Lagrange multipliers.
The model parameters are optimized using AdamW~\cite{Loshchilov22-DecoupledWeight} with a learning rate of $\eta_\mathrm{primal}$, and a weight decay of $\omega_\mathrm{primal}$.
The shared Lagrange multipliers are optimized using AdaMax~\cite{Kingma17-AdamMethodStochastic} with a learning rate of $\eta_\mathrm{sh}$ and a weight decay of $\omega_\mathrm{sh}$.
For the pointwise Lagrange multipliers, we use SGD with a learning rate of $\eta_\mathrm{pw}$ and a weight decay of $\omega_\mathrm{pw}$.
Both supervised and primal-dual models are trained for 5000 epochs for each power system dataset.
This results in five trained models for each power system dataset.

\subsection{Evaluation}\label{sec_evaluation}

After training, each model is evaluated on a held-out test set.
We improve the interpretability of the results by substituting the equality constraints and normalizing the inequality constraints.
Eliminating the equality constraints is performed by solving for the voltages at each bus given the predicted power generation using the PowerModels.JL~\cite{Coffrin18-PowerModelsjl}.
The solved voltages are then used to compute inequality constraint violations.
To normalize the inequality constraint violations, for each element in $\bfg$, we divide by the feasible range of the variable.
After this transformation, constraint violations are on a unit-less, \emph{relative} scale.
Similarly to Equation~\eqref{ch4:eq:inequality_violation}, we can express the \emph{relative} constraint violation as,
\begin{equation}
    \hat{\bm\varepsilon}^g_i = [ \hat{\bfg}(\bfs^d_i, \hat{\bfs}^g_i, \hat{\bfv}_i) ]_+
\end{equation}
where $[\cdot]_+$ computes $\max(0, \cdot)$ for each element of the vector.
The transformed inequality constraints are given by,
\begin{equation}
    \hat{\bfg}(\bfs^d, \bfs^g, \bfv) = \bmat{
        \Re(\bfs^g - \bfs^g_\text{min}) / \Re(\bfs^g_\text{max} - \bfs^g_\text{min})\\
        \Re(\bfs^g_\text{max} - \bfs^g) / \Re(\bfs^g_\text{max} - \bfs^g_\text{min})\\
        \Im(\bfs^g - \bfs^g_\text{min}) / \Im(\bfs^g_\text{max} - \bfs^g_\text{min})\\
        \Im(\bfs^g_\text{max} - \bfs^g) / \Im(\bfs^g_\text{max} - \bfs^g_\text{min})\\
        (|\bfv| - \bfv_\text{min}) / (\bfv_\text{max} - \bfv_\text{min})\\
        (\bfv_\text{min} - |\bfv|) / (\bfv_\text{max} - \bfv_\text{min})\\
        (\bff_f - \bff_{\text{max}}) / (\bff_{\text{max}})\\
        (\bff_t - \bff_{\text{max}}) / (\bff_{\text{max}})\\
        (\angle (\bfv_f \bfv_t^*) - \bm\theta_{\text{min}}) / (\bm\theta_{\text{max}} - \bm\theta_{\text{min}})\\
        (\bm\theta_{\text{min}} - \angle (\bfv_f \bfv_t^*)) / (\bm\theta_{\text{max}} - \bm\theta_{\text{min}})
    }.
\end{equation}
which is a rescaling of the original inequality constraint function $\bfg$ from \eqref{eqn_opf_problem}

Using the above methodology, we compute several metrics for \emph{each sample} in the test set: the optimality gap, the mean violation, and the maximum violation.
The \emph{optimality gap} is the one minus the ratio of the predicted generation cost and the ground truth generation cost.
The \emph{average violation}, equal to $\lVert \hat{\bm\varepsilon}^g_i \rVert_1 / N_g$, is the average of the absolute violations of the equality constraints across all buses and branches, but within each data sample.
Similarly, the \emph{maximum violation}, equal to $\lVert \hat{\bm\varepsilon}^g_i \rVert_\infty$, is the maximum violation of the equality constraints within each data sample.
Note that each metric is computed for each sample in the test set.
Therefore, we report several statistics of the metrics across all samples, including the mean, standard deviation, minimum, and 95th percentile.

\red{

\nocite{Piloto24-CANOSFastScalable, Liu22-TopologyawareGraph,Wachter06-Implementation}

}

\end{document}